# Oxygen and Aluminum-Magnesium Isotopic Systematics of Presolar Nanospinel Grains from CI Chondrite Orgueil

**Nan Liu [1,*], Nicolas Dauphas [2], Sergio Cristallo [3,4], Sara Palmerini[4,5], Maurizio Busso[4,5]**

[1] Department of Physics, Washington University in St. Louis, MO 63130, USA

[2] Origins Laboratory, Department of the Geophysical Sciences and Enrico Fermi Institute, The University of Chicago, IL 60637, USA

[3] INAF, Osservatorio Astronomico d'Abruzzo, Via Mentore Maggini snc, 64100 Teramo, Italy

[4] INFN, Sezione di Perugia, Via A. Pascoli snc, 06123 Perugia, Italy

[5]Department of Physics and Geology, University of Perugia, Via A. Pascoli snc, I-06123 Perugia, Italy

* Correspondence: nliu@physics.wustl.edu

**Abstract:** Presolar oxide grains have been previously divided into several groups (*Group* 1 to 4) based on their isotopic compositions, which can be tied to several stellar sources. Much of available data was acquired on large grains, which may not be fully representative of the presolar grain population present in meteorites. We present here new O isotopic data for 74 small presolar oxide grains (~200 nm in diameter on average) from Orgueil and Al-Mg isotopic systematics for 25 of the grains. Based on data-model comparisons, we show that (*i*) *Group* 1 and *Group* 2 grains more likely originated in low-mass first-ascent (red giant branch; RGB) and/or second-ascent (asymptotic giant branch; AGB) red giant stars and (*ii*) *Group* 1 grains with $(^{26}Al/^{27}Al)_0 \gtrapprox 5\times10^{-3}$ and *Group* 2 grains with $(^{26}Al/^{27}Al)_0 \lessapprox 1\times10^{-2}$ all likely experienced extra circulation processes in their parent low-mass stars but under different conditions, resulting in proton-capture reactions occurring at enhanced temperatures. We do not find any large $^{25}Mg$ excess in *Group* 1 oxide grains with large $^{17}O$ enrichments, which provides evidence that $^{25}Mg$ is not abundantly produced in low-mass stars. We also find that our samples contain a larger proportion of *Group* 4 grains than so far suggested in the literature for larger presolar oxide grains (≥ 400 nm). We also discuss our observations in the light of stellar dust production mechanisms.

## 1. Introduction

Presolar grains can be found in primitive extraterrestrial materials and exhibit extremely large isotopic anomalies, which point to their formation around stars that lived and died prior to the formation of the solar system (see Zinner 2014 for a review). Various presolar phases have been identified, which sample a wide variety of stellar environments including asymptotic giant branch (AGB) stars, core collapse supernovae (CCSNe), and possibly novae (see Zinner 2014 and Nittler & Ciesla 2016 for reviews). The AGB phase represents a stage in the evolution of stars with initial masses of <8 $M_{\odot}$ when they build a CO core (due to exhausted He-burning) and go through a series of instabilities in the envelope, resulting in significant mass loss through stellar winds. CCSNe represent the endpoint of stars more massive than ~10 $M_{\odot}$ and occur as the Fe-rich core, which is left after successive burning stages, cannot burn further to sustain the pressure of the star and experiences neutronization. The star collapses under its own pressure, and the resulting rebound of outer shells onto the Fe core leads to a supernova explosion. Finally, novae are binary systems in which a white dwarf (WD) accretes H and He from a companion main-sequence star, leading to rapid and episodic burning and ejection of the accreted material.

Based on comparisons between isotopic data and models of stellar nucleosynthesis for O isotopes, presolar O-rich grains (silicates, oxides) have been divided into four main groups (Fig. 1), each potentially representing a group of grains from a specific type of star (Nittler & Ciesla 2016). *Group* 1 and *Group* 2 O-rich grains are thought to have come from low-mass ($\lessapprox$2.0 $M_{\odot}$) red giant branch (RGB) and asymptotic giant branch (AGB) stars, in which case the large $^{18}$O depletions of *Group 2* grains could have resulted from enhanced circulation of materials between the convective envelope and H-burning shell (Wasserburg et al. 1995; Nollett et al. 2003; Palmerini et al. 2011). The origin of *Group* 2 grains, however, is equivocal as a recent determination of the proton-capture rate of $^{17}$O also suggests derivation of *Group* 2 grains from intermediate-mass AGB stars (4−8 $M_{\odot}$) (Lugaro et al. 2017). Low-metallicity AGB stars were proposed as the parent stars of *Group* 3 grains, and CCSNe as the parent stars of *Group* 4 grains (Nittler et al. 2008; Nittler & Ciesla 2016). Recently, Nittler et al. (2020) proposed that *Group* 3 characterized by large $^{17}$O depletions were also likely sourced from CCSNe as these *Group* 3 grains could have sampled more $^{16}$O-rich materials than *Group* 4 grains, resulting in their larger $^{16}$O enrichments (These $^{17}$O-depleted *Group* 3 grains will be classified as *Group* 4 grains for discussion hereafter). Grains with extreme $^{17}$O excesses are thought to have come from novae (Nittler et al. 1997; Gyngard et al. 2010).

Ascertaining the stellar origin of presolar O-rich grains, which is solely based on O isotopes, would greatly benefit from using additional isotopic systems. This is because (*i*) stellar models are still incapable of fully capturing all physical processes occurring in stars, *e.g.*, magnetic-buoyancy-induced mixing (Nucci & Busso 2014), so that the reliability of model predictions is uncertain, and (*ii*) stellar nucleosynthesis calculations are also affected by uncertainties in the initial stellar composition and nuclear reaction rates.

Previous studies showed that some *Group* 1 and *Group* 2 O-rich grains exhibit highly variable Mg isotopic compositions (e.g., Gyngard et al. 2010), signatures that are difficult to reconcile with state-of-the-art low-mass RGB/AGB nucleosynthesis model calculations (Cristallo et al. 2009, 2011; Karakas & Lugaro 2016) because the models predict negligible variations in Mg isotopic ratios, other than those inherited from the interstellar medium (ISM) when the star formed. An additional complication arises from the fact that different studies (Zinner et al. 2005; Nittler et al. 2008; Gyngard et al. 2010) reported different Mg isotopic signatures in *Group* 1 and *Group* 2 grains. Large $^{25}$Mg depletions were only observed in the study of Gyngard et al. (2010), calling for further investigation.

The vast majority of presolar silicates, the most abundant type of presolar grain in primitive meteorites and IDPs (up to several hundred ppm, extending to 1.5%), are ferromagnesian silicates (Floss & Haenecour 2016) and suitable for Mg isotopic analysis using secondary ion mass spectrometry (SIMS). Tens of thousands of presolar silicates, ~250 nm in size on average, have been identified by *in situ* O isotopic imaging of primitive fine-grained chondritic matrices mainly using Cameca NanoSIMS instruments. However, fewer than 50 grains were analyzed for Mg isotopes (Presolar Grain Database; Hynes & Gyngard 2009), and the data may have suffered from Mg contamination since the primary $O^-$ ion beam (0.5−1 μm) produced by a duoplasmatron oxygen ion source was often larger than the analyzed grains in size. The challenge in obtaining intrinsic Mg isotopic compositions of presolar silicates is the high spatial resolution required for such analysis with SIMS instruments. This is because presolar silicates cannot be isolated by digesting away solar grains with strong reagents (Oelkers et al. 2018) and they must be identified *in situ* in fine-grained extraterrestrial materials containing abundant solar Mg-rich minerals. This can easily lead to sampling Mg signals from adjacent solar Mg-rich grains during the analysis, if the primary $O^-$ beam is larger than the grain in size. Previous studies attempted to isolate targeted presolar silicate grains using a focused ion beam instrument prior to Mg isotope analysis using Duoplasmatron-NanoSIMS (Nguyen & Messenger 2014; Kodolányi et al. 2014), based on which Kodolányi et al. (2014) reported small

$^{25,26}$Mg excesses (within ~10% of the solar abundances) for *Group* 1 silicate grains. The recently developed Hyperion radio-frequency ion source can produce a ~100 nm primary $O^{-}$ ion beam of a few pA on the NanoSIMS (Malherbe et al. 2016). This makes it possible to better measure the undiluted Mg isotopic compositions of presolar silicates. Magnesium isotopic analysis with a Hyperion source on NanoSIMS instruments revealed large $^{25}$Mg excesses (several times of the solar abundance) in a number of *Group* 1 and *Group* 2 presolar silicate grains (Leitner & Hoppe 2019; Verdier-Paoletti et al. 2019; Leitner et al. 2019), which are again incompatible with state-of-the-art nucleosynthesis model predictions for low-mass AGB stars of different metallicities (Cristallo 2009, 2011; Karakas & Lugaro 2016) and thus pose a challenge to their proposed low-mass stellar origins. Intermediate-mass AGB stars, massive stars, and CCSNe have been proposed as stellar sources of $^{25}$Mg-rich *Group* 1 grains (Leitner & Hoppe 2019; Verdier-Paoletti et al. 2019; Hoppe et al. 2021).

In addition to presolar silicates, presolar spinel ($MgAl_2O_4$) grains also carry a significant amount of Mg and are, therefore, suitable for investigating Mg isotopic ratios. In contrast to presolar silicates, presolar spinel grains can be extracted by acid dissolution of primitive meteorites and concentrated in acid residues. Around 60 spinel grains were analyzed for their Mg isotopes in previous studies (Zinner et al. 2005; Nittler et al. 2008; Gyngard et al. 2010), which pointed to close-to-solar $^{25}$Mg/$^{24}$Mg in *Group* 1 and *Group* 2 spinel grains. In contrast, large $^{25}$Mg excesses (*i.e.*, >15% higher than solar $^{25}$Mg/$^{24}$Mg) were observed in 16 of 97 (16%) *Group* 1 presolar silicates (Hoppe et al. 2018, 2021; Leitner & Hoppe 2019). The inferred percentage of $^{25}$Mg-rich *Group* 1 silicate grains, 16%, likely represents a lower limit due to, *e.g.*, the likelihood of mislocating targeted O-anomalous grains during Mg isotopic analyses (Hoppe et al. 2021). For comparison, none of the investigated *Group* 1 spinel grains (Zinner et al. 2005; Nittler et al. 2008; Gyngard et al. 2010) had $^{25}$Mg/$^{24}$Mg >15% higher than the solar value. The different Mg isotopic compositions observed between *Group* 1 spinel and silicate grains could reflect differences in their stellar origins. Hoppe et al. (2021) showed that *Group* 1 silicate grains with high $^{17}$O/$^{16}$O ratios are more likely to exhibit large $^{25}$Mg excesses: >15% higher-than-solar $^{25}$Mg/$^{24}$Mg ratios were found in only 16% of all *Group* 1 silicates but in 30% of those with $^{17}$O/$^{16}$O ratios above $1.0 \times 10^{-3}$. Given that only a few *Group* 1 spinel grains previously studied had such high $^{17}$O/$^{16}$O ratios, it is desirable to analyze more *Group* 1 spinel grains with high $^{17}$O/$^{16}$O ratios to investigate whether there is any systematic difference in the Mg isotopic signature between *Group* 1 spinel and silicate grains.

For direct comparison with the recent high-spatial-resolution Mg isotopic data for presolar silicates (Hoppe et al. 2018, 2021; Leitner & Hoppe 2019), we investigated Al-Mg isotopic systematics of presolar oxide grains identified in the spinel-rich acid residue of the Orgueil CI chondrite, using Hyperion-NanoSIMS with a focus on grains with high $^{17}O/^{16}O$ ratios. The obtained Al/Mg ratios allow us to assess the effect of $^{26}Al$ decay ($t_{1/2}$ = 0.72 Ma) on the $^{26}Mg$ budget for each of the grains. Here we report O isotopic data for 74 new presolar oxides and Mg isotopic data (and inferred initial $^{26}Al/^{27}Al$ when available) for 25 of the grains (24 spinel grains and one Al-rich oxide), including one putative nova, four *Group* 2, and 20 *Group* 1 grains.

## 2. Materials and Methods

### *2.1. NanoSIMS Isotopic Analyses*

The analyzed grains were all found on the < 200 nm spinel-rich Orgueil separate mount previously prepared and studied by Dauphas et al. (2010). We imaged ~140 15 × 15 μm$^2$ areas (256 × 256 pixels) for O isotopes using a ~1 pA, 16 keV $Cs^+$ beam on the NanoSIMS 50 instrument at Washington University in St. Louis. We used entrance slit #3 and aperture slit #2 and put in the energy slit to cut off 20% of the $^{16}O^-$ signal. Our achieved mass resolving power (MRP) resulted in <1% contribution from $^{16}OH^-$ to $^{17}O^-$, which was estimated based on high-resolution mass scans of O-rich particles on the sample mount and by assuming a symmetric $^{16}OH^-$ peak. We collected $^{16}O^-$, $^{17}O^-$, $^{18}O^-$, $^{28}Si^-$, and $^{27}Al^{16}O^-$ secondary ions simultaneously on electron multipliers (EMs) in the multicollection mode for about two hours (100 ion images) for each area.

We then relocated 35 Al-rich grains mostly with high $^{17}O/^{16}O$ ratios by comparing the previously collected $^{27}Al^{16}O^-$ and $^{28}Si^-$ images with $^{27}Al^+$ and $^{28}Si^+$ images for the Al-Mg isotopic analyses (Fig. 2). The ratios needed for application of $^{26}Al$-$^{26}Mg$ systematics ($t_{1/2}$=0.7 Ma) are $^{25}Mg/^{24}Mg$, $^{26}Mg/^{24}Mg$, and $^{27}Al/^{24}Mg$. These ratios were analyzed by rastering a ~1 pA, 16 keV primary $O^-$ ion beam produced by a Hyperion RF source over 5 × 5 μm$^2$ areas (256 × 256 pixels). The source was operated at a frequency of 40 MHz, a power of 800 W, and a voltage of −8 kV. We used the same set of slits, and our mass-resolving power resulted in <0.1% contributions from $^{24}MgH^+$ and $^{25}MgH^+$ signals on $^{25}Mg^+$ and $^{26}Mg^+$ signals, respectively. We collected $^{24}Mg^+$, $^{25}Mg^+$, $^{26}Mg^+$, $^{27}Al^+$, and $^{28}Si^+$ secondary ions simultaneously on EMs in multicollection mode; the analyses continued until the Mg signals dropped significantly. Reliable Mg isotope ratios were obtained in 25 out of the 35 selected grains. For the other 10 grains, there was insufficient Mg signal to obtain precise Mg isotopic data or Mg contamination from adjacent areas to the targeted grains was significant. NIST glass reference material SRM 610 was measured for determining the Mg/Al relative sensitivity factor (RSF), $(Mg/Al)_{tru} = (Mg/Al)_{meas}/RSF$, in which $(Mg/Al)_{tru}$ and $(Mg/Al)_{meas}$ are the true and measured Mg/Al ratios, respectively. The Mg/Al RSF was determined to be 0.83 with ±15% 1σ uncertainties.

In both negative ($O^-$ ion source for $^{26}$Al-$^{26}$Mg systematics) and positive ($Cs^+$ ion source for O isotopic analyses) NanoSIMS modes, we were able to achieve a spatial resolution of ~100 nm, which, given the low grain-density of our sample mount (Fig. 3a), was sufficient to effectively suppress contamination from adjacent grains for the 25 grains as illustrated in Fig. 2. In other words, our obtained isotopic data were not significantly affected by isotopic dilution and represent the intrinsic isotopic compositions of the grains. All the isotopic data were collected in imaging mode and reduced using the IDL (L3 Harris Geospatial Solutions, Inc.) based *L'image* software package (Version: 12-16-2020; Nittler et al. 2018).

### *2.2. Isotopic Data Reduction Using L'image Software*

Figure 3 illustrates the procedure for reducing the O isotopic data using *L'image*. The mean O isotopic ratios of each analyzed area (with the O-anomalous grains excluded) were used as the standard values for normalization. In σ plots (panels d and f in Fig. 3), each pixel represents the number of standard deviations that its measured isotopic ratio is away from the mean isotopic ratio of the whole area. For each area, we calculated the O isotopic compositions of all O-rich particles (e.g., panels g-h in Fig. 3), which were automatically identified using the auto-particle identification function (with parameters adjusted to maximally capture O-rich grains in the $^{16}O^-$ image for each area) in the *L'image* software.

Out of ~30,000 O-rich grains (with a median diameter of 290 nm) identified in the ion images, we detected 74 outliers (0.25% of all the O-rich grains, with a median diameter of 300 nm). Our procedure for detecting presolar grains uses the ellipse envelope and local outlier factor algorithms (see supplementary online materials for details). The ellipse envelope algorithm assumes that the data set free of outliers would follow a bivariate normal distribution in $\delta^{17}O$-$\delta^{18}O$, an assumption that we validated based on analyses of carbonaceous chondritic matrices free of presolar O-rich grains due to extensive aqueous alteration (Liu et al. 2020). We further tested the result of the ellipse envelope algorithm by applying the local outlier factor algorithm, which considers as outliers the grains that have a substantially lower density than their neighbors. We found a good agreement in the detected outliers between the two algorithms. We also excluded grains with <15% deviations from the terrestrial O isotopic composition, given the range of O isotopic variations observed in the solar system (Sakamoto et al. 2007; McKeegan et al. 2011). Among the 74 identified presolar grains, 72 grains are Al-rich (AlO hotspots), and two grains are Al-poor ($AlO^-$ signal enrichments, i.e., comparable to adjacent background signals) according to the $^{27}Al^{16}O^-$ ion images. The inferred proportion of our presolar O-rich grains, 0.25%, is significantly higher than the percentage of outliers beyond 4σ (0.0055%) expected for a 2D-Gaussian distribution (Wang et al. 2015), thus demonstrating that our identified presolar grains are statistically significant. We also confirmed the O isotopic anomalies of all the 74 identified presolar grains by manually examining the isotopic images of each analyzed area (e.g., panels c-f in Fig. 3).

The final O and Mg isotopic data (Tables 1 and A1) were further analyzed by setting small regions of interest (ROIs) to exclude extrinsic O and Mg signals from adjacent particles. We observed some long-term (on the order of several days) drift in the O isotopic ratios, mainly caused by the aging of the EM used for counting $^{16}O$ over time. Given the EM aging problem, it is more appropriate to use the O isotopic ratios of the area where a grain is located for normalization compared to the mean of all the analyzed areas (with the O-anomalous grain excluded). Thus, the O isotopic data were normalized to the mean of the corresponding area, and the 1σ systematic errors for O isotopes were estimated based on our observed daily standard deviations (~2% in the $^{17}O/^{16}O$ ratio and ~1% in the $^{18}O/^{16}O$ ratio). Since the Mg count rates were much lower and the data were collected within a week, the Mg isotopic data were simply normalized to the mean of the 25 areas where the grains were located (with the O-anomalous grain excluded), and the 1σ systematic errors for Mg isotopes were estimated based on the standard deviations of the 25 areas (19‰ and 18‰ 1 σ errors in $\delta^{25}Mg$ and $\delta^{26}Mg$, respectively). The 1σ errors for O and Mg isotopic ratios reported in Tables 1 and A1 include both and the systematic errors and Poisson errors.

All the O isotopic data were automatically corrected for the effects of dead time and quasi-simultaneous arrivals (QSA) in the *L'image* (see Nittler et al. 2021 for details). The QSA effect on the $^{16}O^-$ image was corrected by setting the parameters *primary current* to 1 pA and *beta* to 0.7 (Jones et al. 2017). Because of our low $^{16}O^-$ count rates (on the order of 10,000 counts/s), the QSA correction was small (<a few ‰). In addition, given our low $^{24}Mg^+$ count rates (a few hundred counts/s), no QSA correction was needed. The Mg/Al ratios were corrected for the RSF, 0.83 ± 0.12 (1σ error). The O and Mg isotopic data of the 25 presolar oxides are reported in Table 1, and the O isotopic data of the other identified presolar oxides are given in Table A1. The delta notation is defined as $\delta^{i}A = [(^{i}A/^{j}A)_{samp}/(^{i}A/^{j}A)_{std}-1] \times 1000$, in which A denotes an element, $^{i}A$ and $^{j}A$ denote two different isotopes of the element A, $(^{i}A/^{j}A)_{samp}$ the isotopic ratio measured in a sample, and $(^{i}A/^{j}A)_{std}$ the isotopic ratio measured in the standard as defined above. The normalizing isotopes used for calculating $\delta^{i}O$ and $\delta^{i}Mg$ values are $^{16}O$ and $^{24}Mg$, respectively.

## 3. Results

### *3.1. Grain size estimate*

Our NanoSIMS ion images suggest that O-rich grains on the mount had a median diameter of 290 nm, which is larger than the grain size estimate (150–200 nm) based on the transmission electron microscope (TEM) observation of grains on the same mount (Dauphas et al. 2010). Given the ~100 nm spatial resolution in our NanoSIMS analyses, our larger grain size estimate could be due to the poorer spatial resolution of the NanoSIMS compared to the nm-scale resolution of the TEM analyses. In principle, we could estimate the true grain size by subtracting a beam diameter in quadrature from the measured diameter (Nittler et al. 2018), which would yield a grain diameter of 272 nm for an original size of 290 nm and a beam diameter of 100 nm. The beam diameter was estimated using the Resolution Calculation in *L'image*, which assumes a Gaussian-shaped beam across a perfectly straight, infinitely long edge and estimates the resolution from the 16-84% levels across the edge. The inferred correction factor based on coordinated Auger and NanoSIMS analyses of presolar silicates (Zhao et al. 2013), however, is two-to-three times larger than calculated using the beam-size-correction equation (Nittler et al. 2018). By adopting the 30% correction as reported by Zhao et al. (2013), it yields a true median grain size of 210 nm for our 74 presolar spinel grains, which is close to the TEM results of Dauphas et al. (2010); this corrected median grain size will be adopted hereafter when comparing to the sizes of literature spinel grains. It is, however, noteworthy that the uncorrected grain sizes reported in Table 1 were estimated in the same way as for presolar silicates in most of previous studies.

### *3.2. Oxygen Isotopic Compositions*

Figure 1 compares the 74 presolar oxides from this study with 404 presolar oxides from the literature (Presolar Grain Database; Hynes & Gyngard 2009) for their O isotopic ratios and groupings (i.e., proportions of each group; panels b-c in Fig. 1). We compare our oxide grain data to the literature data for presolar oxides (mostly from Choi et al. 1998, 1999; Gyngard et al. 2010; Krestina et al. 2002; Nguyen et al. 2003; Nittler et al. 2008; Zinner et al. 2003, 2005), which consist mainly of larger oxide grains (generally >0.4 μm to a few μm in size) and thus allow us to investigate whether there is any size dependence on the isotopic compositions of the grains and their statistical distributions. We excluded (*i*) the small presolar oxides identified in Murray CF acid residue (a mean size of 150 nm; Nguyen et al. 2003; Zinner et al. 2003) and in *in situ* measurements of chondritic matrices (<30 presolar oxide grains compiled in the PGD), and (*ii*) the presolar oxides identified solely based on $^{18}O/^{16}O$ in the studies of Nittler et al. (1994, 1997) as their population distribution is biased (*i.e.*, grains solely with $^{17}O$ anomalies could not be identified in these studies). We adopted the revised scheme proposed by Nittler et al. (2020) for classifying grains examined in this study and the literature: Nova grains are those that have $^{17}O/^{16}O \geq 0.005$, *Group* 1 grains have $4.56 \times 10^{-4} \leq {}^{17}O/^{16}O < 0.005$ and $0.001 \leq {}^{18}O/^{16}O < 0.0024$, *Group* 2 grains have $^{18}O/^{16}O < 0.001$ with higher-than-terrestrial $^{17}O/^{16}O$ ratios, *Group* 3 grains have lower-than-terrestrial $^{17}O/^{16}O$, $^{18}O/^{16}O$ and $^{18}O/^{17}O$ ratios, and *Group* 4 grains that either have $^{17}O/^{16}O \leq 3.04 \times 10^{-4}$ with higher-than-terrestrial $^{18}O/^{17}O$ ratios or have $^{18}O/^{16}O$ ratios $\geq 0.0024$ with $^{18}O/^{17}O$ ratios > 3.

Figure 1a shows that our *Group* 1, *Group* 3, and nova grains have O isotopic compositions like literature oxide data. However, compared to the literature data, our *Group* 2 grains have smaller $^{18}O$ depletions, and some of our *Group* 4 grains have $^{17}O$ depletions, which were previously found only in very few spinel grains. The smaller $^{18}O$ depletions of our *Group* 2 grains are likely caused by contamination of grain surfaces by materials with normal O isotopic composition. For example, mixing $^{18}O$-depleted stellar materials with 10% solar material can enhance the $^{18}O/^{16}O$ ratio from $\sim 1 \times 10^{-7}$ to $(3–4) \times 10^{-4}$ (Lugaro et al. 2017). Previously reported large $^{18}O$ depletions were mainly found in large oxide grains that were a few μm in size and had negligible surface O contamination (e.g., Nittler et al. 1997). While $^{17}O$-depleted *Group* 4 oxide grains are expected to be rare according to previous studies of large oxide grains (0.02%, one out of the 404 grains from the literature; Fig. 1), nine out of our 15 *Group* 4 oxides (12.2% among all the 74 grains) are $^{17}O$-depleted. For comparison with the O isotopic data of our small oxide grains, Nguyen et al. (2003) and Zinner et al. (2003) also investigated the O isotopic compositions of oxides (mostly spinel) in the Murray CF acid residue (with a mean grain size of 0.15 μm) and reported a proportion of 0.2% for presolar spinel grains, similar to our result (0.25%; see Section 2.2). However, their CF presolar oxide grains had very restricted O isotopic compositions, *i.e.*, closer-to-solar O isotopic compositions, and almost all of the grains belonged to *Group* 1, in contrast to the more anomalous and diverse O isotopic compositions of larger presolar oxides from Murray CG acid residue (of a mean size of 0.45 μm) found in the same studies. The authors ascribed the restricted O isotopic compositions of CF presolar oxides to isotopic dilution due to their small sizes. In contrast to these previous studies, our grains, despite the small grain sizes (~200 nm on average after correction), exhibit variable O isotopic signatures that are generally agree with those of larger oxide grains as shown in Fig. 1a, thus pointing to less severe O isotopic dilution during our grain analysis (Figs. 2, 3). Compared to the study of Nguyen et al. (2003), our less diluted O isotopic data likely result from both the lower grain density on our sample mount (Fig. 3) and the better spatial resolution achieved in our study (1 pA $Cs^+$ beam at 59 nm/pixel resolution in this study vs. 6 pA $Cs^+$ beam mostly at 78 nm/pixel resolution in Nguyen et al. 2003).

Examination of the grouping distribution (Figs. 1b,c) shows that our presolar oxide grains comprise a higher proportion of *Group* 4 grains than literature data for larger presolar oxide grains (20.3% vs. 7.2%). The *Group* 4 grain abundance from this study is $20.3^{+14.7}_{-11.2}$ (2σ errors; Gehrels 1986). For comparison, the literature data yield a *Group* 4 grain abundance of $7.2^{+3.4}_{-2.4}$. Given that *Group* 4 grains were inferred to have come from CCSNe (e.g., Nittler et al. 2020), our result suggests that compared to large presolar spinel grains reported in the literature, our small presolar spinel grains (~200 nm on average) seem to have sampled a higher abundance of CCSN grains (about 2σ significance), in line with the previous observation of enhanced abundances of *Group* 3 and *Group* 4 silicate grains in smaller size fractions (Hoppe et al. 2015).

### *3.3. Aluminum-magnesium isotopic systematics*

The TEM-EDX measurements of Dauphas et al. (2010) revealed that the acid residue on our sample mount consists dominantly of spinel grains. The Al/Mg elemental ratios (Table 1, determined from the NanoSIMS analyses after RSF correction) show that among the 25 O-anomalous grains, 24 grains had Al/Mg ratios close to the stoichiometric Al/Mg value (atomic Al/Mg = 2; $MgAl_2O_4$) of spinel and are thus probably $MgAl_2O_4$; one grain had an Al/Mg ratio of ~150 and is likely corundum (nominally $Al_2O_3$). Figure 4 shows that the O-anomalous Al-rich grains have higher Al/Mg ratios than the solar Al-rich grains present in the same Al-Mg ion images. Based on coordinated NanoSIMS, SEM-EDX, and TEM-EDX analyses, Liu et al. (2021) obtained consistent Al/Mg RSF values for Burma spinel and SiC, which have vastly different chemistries, and Liu et al. (2018a) also did not find any measurable difference in the determined Al/Mg RSF value inferred from NanoSIMS analyses of SRM 610 glass and Burma spinel. Our derived Al/Mg ratios for solar Orgueil spinel grains (gray histograms in Fig. 4) are in general agreement with those for solar spinel grains reported by Dauphas et al. (2010) based on TEM-EDX analyses, supporting the accuracy of our data. The systematic difference in chemical composition between anomalous and normal spinels is, therefore, likely to be real. Such a difference in the Al/Mg ratio between presolar and solar Al-rich grains in the spinel-rich acid residues of chondrites was observed in previous NanoSIMS studies (Zinner et al. 2005; Gyngard et al. 2010). However, Gyngard et al. (2010) raised concerns regarding this observation, as in previous studies (*i*) spinel clumps instead of single spinel grains were used for determining the RSF, but some previous NanoSIMS studies have hinted at elemental ratio variations as a function of grain size, (*ii*) the clumps of grains used for normalization could have contained grains of non-spinel ($MgAl_2O_4$) composition inherited from the residue itself, thereby compromising the determination of the Al/Mg RSF value, and (*iii*) the $O^-$ beam was large in size (0.5−1 μm), and they could not absolutely rule out the possibility of sampling Al contamination during the presolar spinel measurements, which could have elevated the measured Al/Mg ratios. In comparison, our $O^-$ beam was much smaller (~100 nm), and the identified Al-rich grains were similar in size and were separated grains instead of grain clumps in most cases. We, therefore, conclude that the different range of Al/Mg ratios observed between presolar and solar spinel grains is likely a real feature.

Figure 5 compares presolar spinel grains from this study with those from the literature (Zinner et al. 2005; Nittler et al. 2008; Gyngard et al. 2010) for their O and Mg isotopic compositions. Note that our systematic errors are only ~20‰ in both $\delta^{25}$Mg and $\delta^{26}$Mg (see Section 2.2 for details) and that uncertainties in our Mg isotopic data are dominated by Poisson statistical errors due to the limited number of Mg counts collected in each grain (~200 nm on average). For comparison, the spinel grains from the literature were larger (~400−700 nm in size) by at least a factor of eight in volume, leading to smaller statistical uncertainties and, consequently, smaller errors in the reduced data. Most of our *Group* 1 grains have close-to-solar Mg isotopic compositions, in agreement with literature data for *Group* 1 grains. In detail, Mg isotopic anomalies above 150‰ were observed in three out of our 19 *Group* 1 spinel grains (15.8%), and eight out of 35 *Group* 1 grains (22.9%) from the literature. In contrast to the large $^{26}$Mg excesses (≥400‰) previously observed in Group 2 grains, we did not find any $^{26}$Mg excesses above 150‰ in the four *Group* 2 grains from this study. All six *Group* 1 and *Group* 2 grains with large $^{25}$Mg depletions and $^{26}$Mg excesses were found in the study of Gyngard et al. (2010). Zinner et al. (2005) did not find any grain with these characteristics. The one nova grain reported in Gyngard et al. (2010) had large excesses in both $^{25}Mg$ and $^{26}Mg$, while our nova grain had only a minor $^{25}Mg$ excess (70 ± 46 ‰, 1σ error). Finally, one of the most $^{17}O$-rich *Group* 1 spinel grains, A20-03-#030, shows a significant depletion in $^{25}Mg$ (−318 ± 110 ‰, 1σ error), which was seldomly seen in *Group* 1 grains with such large $^{17}O$ enrichment but is comparable to the Mg isotopic signature of *Group* 1 silicate grain MET_01B_53_1 ($^{17}O/^{16}O=10.06\pm0.95\times10^{-4}$) reported by Hoppe et al. (2021).

As shown in Fig. 5a, our *Group* 2 grains are less depleted of $^{18}O$ than those from the literature. Contamination with isotopically normal oxygen could have taken place in the nebula by gas-solid interaction or on the parent-body through aqueous alteration. We are unable to quantify and correct the extent to which such surface-seated O contamination could have affected $^{18}O/^{16}O$ ratios in our *Group 2* grains. Larger presolar oxide groups are expected to be less affected by such surface O-contamination because of their lowered surface-to-volume ratios.

## 4. Discussion

### *4.1. Mg Isotope Exchange in Presolar Spinel?*

Nittler et al. (2008) suggested that Mg isotopes in presolar spinel grains may have experienced exchange with a gaseous reservoir in the ISM, thereby modifying their compositions to a more average ISM composition. The proposal was based on the observations that (*i*) many of the presolar spinel grains studied by Zinner et al. (2005) and Nittler et al. (2008) had terrestrial Mg isotopic composition while the presolar hibonite/corundum grains from Nittler et al. (2008) seemed more anomalous in their Mg isotopic compositions and (*ii*) some calcium-aluminum-rich (CAI) inclusions in meteorites tend to show more significant Mg isotopic fractionation in the interior than at the surface, indicating equilibration with a gaseous reservoir for Mg isotopes but not for O isotopes (e.g., Simon et al. 2005). In the ISM-exchange scenario, it is unclear how the proposed Mg equilibration could have operated, because, unlike the high-density (*e.g.*, $10^{-5}$ atm), high-temperature region (e.g., ~1000–2000 K) in the solar nebula where CAI minerals could have condensed and been processed (Macpherson et al. 2005), the gas in the ISM is either cool/dense ($10^6$ molecules per $cm^3$, 10–20 K) or hot/diffuse ($10^{-4}$ ions per $cm^3$, $10^6$–$10^7$ K) (Ferrière 2001), which would not be conducive to isotope exchange between dust and gas.

As detailed below, the different Al/Mg ratios of our solar and presolar spinel grains provide a hint that the presolar grains analyzed did not experience significant Mg isotope equilibration in the solar nebula. The TEM-EDX data reported in Dauphas et al. (2010) revealed a wide range of Al/Mg ratios for Al-rich spinel grains on the same sample mount studied here (Fig. 4), mostly due to variable amounts of Cr substituting for Al in the Mg, Al-rich end member of the spinel group, $MgAl_2O_4$. This explains why the Al/Mg ratio is commonly observed to lie below two in the solar spinel grains, which are common minerals of CAIs and chondrules, identified in our ion images (Fig. 4). The wide range of Al/Mg ratios observed in our solar spinel grains likely reflects their varying thermal alteration histories in the early solar nebula, as supported by the reduced Mg isotopic fractionation toward the surface observed in CAIs (e.g., Simon et al. 2005). In comparison to solar Orgueil spinel grains, our presolar spinel grains are more tightly grouped around Al/Mg = 2, i.e., more similar to $MgAl_2O_4$ in composition. Thus, it is likely that presolar spinel grains were not transported to high-temperature nebular regions where high-temperature meteoritic components, CAIs and chondrules, were present; otherwise, we would observe similar, lowered Al/Mg ratios in presolar spinel grains due to Cr substitution. Note that we do not expect the envelope compositions of the parent stars of *Group* 1 and *Group* 2 grains to differ significantly from the solar composition as these grains are mainly sourced from low-mass, close-to-solar-metallicity RGB/AGB stars as will be discussed in Section 4.2.

In the following, we assume that the Mg isotopic data of presolar spinel grains from this study represent the intrinsic composition of their parent stars for discussion.

### *4.2. Presolar Spinel Grains from RGB/AGB Stars*

Figure 5 summarizes the known effects of different astrophysical processes on O and Mg isotopic ratios. We refer the reader to Nittler et al. (2008) for a detailed overview of RGB/AGB stellar nucleosynthesis, Galactic chemical evolution (GCE), and nonstandard stellar mixing processes in the context of O and Mg isotopes. In the following sections, we will focus our discussion on the stellar origins of presolar O-rich grains in the context of AGB and nova nucleosynthesis and GCE. GCE describes the process by which the elemental and isotopic composition of the Galaxy varies in time and place because of stellar nucleosynthesis and material cycling between stars and the ISM (Nittler & Dauphas 2006). The isotopes $^{16}O$ and $^{24}Mg$ are primary, meaning that they can be made in the first generation of stars, while isotopes $^{17}O$, $^{18}O$, $^{25}Mg$, and $^{26}Mg$ are secondary, meaning that their production requires the preexistence of primary isotopes. As a result, the abundances of secondary isotopes increase with increasing Galactic age, i.e., increasing metallicity of ISM gas (Timmes et al. 1995). In a 3-isotope plot (in delta notation) for O, Mg, and Si isotopes simple GCE models would predict that the data should plot roughly along a slope-1 line (e.g., Meyer et al. 2008). More sophisticated GCE models, however, yield slopes that can depart from one (Gaidos et al. 2009 predict a slope of ~0.5 for O isotopes).

*4.2.1. RGB/AGB Stars and Group 1 spinel grains*

*4.2.1.1. RGB/AGB stars*

We provide a brief overview of the RGB/AGB phase here to help the reader understand the RGB/AGB models presented in the following sections. We refer the reader to Busso et al. (1999), Herwig (2005), Straniero et al. (2006), and Karakas & Lattanzio (2014) for more details regarding RGB/AGB stellar evolution. When a low- or intermediate-mass star ($\lesssim$8 $M_{\odot}$) runs out of H fuel in its core, it expands and cools, becoming a red giant that is located on the RGB on a Hertzsprung-Russell diagram. Shortly after ascending the RGB, the star undergoes an episode of deep convection, known as the first dredge-up (FDU), which changes the stellar surface composition by mixing the partial H-burning products from deep layers of the star into the envelope. Following core He-burning after the RGB phase, the star expands again and ascends the AGB. An AGB star consists of a partially degenerate CO core, a He-burning shell, a He-intershell, a H-burning shell, and a large convective H-rich envelope. The energy necessary to sustain the surface luminosity is provided by the H-burning shell, which is recurrently turned off by a sudden activation of the He-burning shell. This is because the He-burning is subject to recurrent thin-shell instabilities, where the released energy is large enough to trigger a dynamic runaway; this last phenonmenon is called a thermal pulse (TP), and the evolutionary stage during which recurrent TPs take place is known as the TP-AGB phase. The occurrence of TPs causes the He-intershell, a thin region ($10^{-2}$–$10^{-3}$ $M_{\odot}$) lying between the H-burning and He-burning layers, to convect and to enrich the He-intershell with $^{12}$C. If the TP is strong enough to turn off H-burning in the H-shell, the convective envelope may penetrate into the He-intershell, bringing the fresh nucleosynthetic products, including $^{12}$C and slow neutron-capture process (*s*-process) nuclides, to the surface. This recurrent mixing episode is known as third dredge-up (TDU). With repeated TDU episodes, the surface C/O ratio increases to above unity for low-mass ($\lesssim$3 $M_{\odot}$) AGB stars. In intermediate-mass AGB stars ($\gtrsim$3 $M_{\odot}$), their stellar envelopes remain O-rich during the AGB phase because of increased dilution and decreased TDU efficency compared to low-mass stars (in particular at solar-like metallicities). Dust grains are expected to condense in the cooling gas that is lost from the stellar surface (Lodders & Fegley 1995; Höfner & Olofsson 2018; Bladh et al. 2019), and the mass loss rate is predicted to increase with the stellar evolution phase. As an example, for a 2 $M_\odot$, 1.4 $Z_\odot$ star (FRUITY calculations; Cristallo et al. 2011), stellar winds account for 3% of the total mass lost by the star on the RGB and up to 65% of that at the tip of the AGB. Thus, the contribution of low-and intermediate-mass stars to the solar system presolar grain inventory during the RGB phase is expected to be significantly less important than that during the AGB phase.

*4.2.1.2. O isotopes*

Stellar nucleosynthesis calculations performed in traditional evolutionary codes show that the $^{17}O/^{16}O$ ratio in low-mass AGB stars derives primarily from the FDU that occurs prior to the AGB phase when the star is on the RGB. The model prediction for $^{17}O/^{16}O$ depends strongly on the stellar mass, with only a minor dependence on the initial composition (*e.g.*, Straniero et al. 2017). This is so because, in the absence of complications like those introduced by partial mixing, in the H-burning regions of low-mass RGB and/or AGB stars the $^{17}O/^{16}O$ ratio is simply given by the inverse ratio of the respective proton-capture reaction rates, $\frac{N(^{17}O)}{N(^{16}O)} = \frac{<\sigma v>[^{16}O(p,\gamma)^{17}F]}{<\sigma v>[^{17}O(p,\gamma)^{18}F]}$, in which σ is the nuclear cross section, ν is relative velocity of the interacting nuclei, and $< \sigma v >$ the Maxwellian-averaged cross section. Straniero et al. (2017) showed that, in such a case, the $^{17}O/^{16}O$ ratio traces the temperature profile within the H-burning zone. The $^{17}O/^{16}O$ ratio thus increases with increasing initial stellar mass, until it reaches the maximum ratio at ~2.5 $M_\odot$. A comparison with the FRUITY AGB model calculations (Cristallo et al. 2009, 2011) in Fig. 6a for $^{17}O/^{16}O$ suggests that all the spinel grains analyzed previously (Zinner et al. 2005; Nittler et al. 2008; Gyngard et al. 2010) came from RGB/AGB stars of ≤1.5 $M_\odot$, while our new spinel grains came from stars of up to 2.0 $M_\odot$. Note that the grain data should be compared to the final composition predicted for each star for $^{17}O/^{16}O$, because (*i*) the stellar surface composition is quickly modified from the initial composition to the final $^{17}O/^{16}O$ ratio after the FDU (*i.e.*, $^{17}O/^{16}O$ barely changes after the FDU in the stellar envelope if the star does not undergo extra mixing) and (*ii*) most of the dust is expected to condense after the last several TPs during the AGB phase (see discussion above).

We further compare our *Group* 1 O-rich grain data with intermediate-mass stellar model predictions in Fig. 7 to explore whether *Group* 1 grains could have been sourced from intermediate-mass stars. The 4.5, 5, and 6 $M_\odot$ models in Fig. 7 were computed using the same code as for the low-mass FRUITY stellar models shown in Fig. 6. A difference, however, compared to the low-mass FRUIT stellar models is that these intermediate-mass stellar models use upgraded physical inputs as described in Vescovi et al. (2020). Specifically, we adopted an improved Equation of State (EoS), which led to higher temperatures at the base of the convective envelope and thus a more efficient activation of hot bottom burning (HBB). HBB refers to the circumstance where the bottom of the convective envelope of an intermediate-mass star reaches temperatures sufficiently high for efficient proton-capture reactions to take place. These HBB models have been presented by Palmerini et al. (2021) for comparison with presolar O-rich grains. Many poorly constrained model parameters affect the evolution of these more massive AGB stars. In the computation, we adopted an extremely reduced mass-loss rate to increase the number of TPs and thereby maximize the effects of HBB. Thus, the intermediate-mass stellar model results shown in Figs. 7b,d need to be considered with the caveat that they are not representative of the diverse outcomes allowed by model uncertainties. A comparison of our HBB models with those presented by Lugaro et al. (2017) points to similar predictions for the final O isotopic compositions, but these authors reported lower final $^{26}Al/^{27}Al$ ratios (e.g., $1.7 \times 10^{-2}$ in our 4.5 $M_{\odot}$ model and $1.7 \times 10^{-3}$ in the 4.5 $M_{\odot}$ model by Lugaro et al. 2017). Our higher $^{26}Al/^{27}Al$ predictions likely reflect the higher H-burning temperatures achieved in our model resulting from our updated EoS and adopted low mass-loss rate. FRUITY model calculations for intermediate-mass AGB stars (Fig. 7b) provide a poor match to the O isotopic compositions of *Group* 1 grains, thus implying that our *Group* 1 spinel grains more likely came from lower-mass AGB stars as shown in Fig. 6. The poor match of intermediate-mass AGB stellar models to *Group* 1 grain data, is also supported by the intermediate-mass AGB models of Lugaro et al. (2017).

We recall that it has been known for more than 30 years (Gilroy & Brown 1991) that most low-mass stars evolving along the RGB are inferred to have experienced non-convective mixing phenomena that significantly alter the C, N, and O isotopic ratios of low-mass stars (Busso et al. 1999). The occurrence of non-convective mixing in RGB stars was suggested by Abia et al. (2012) to account for the C and O isotopic ratios observed in two well-known low-mass red giant stars (α Boo and α Tau). These poorly understood mixing processes (also known as extra mixing on the RGB) were attributed over the years either to unknown slow circulation effects (named cool bottom processing in parameterized models; see *e.g.*, Nollett et al. 2003, Palmerini et al. 2011) or to the onset of various kinds of diffusive mixing (Eggleton et al. 2006). More recently, these extra mixing processes were also proposed to result from buoyancy-induced circulation in magnetically active low-mass stars (Palmerini et al. 2017), in which case the induced mixing is expected to extend to the AGB phase. For the scopes of the present work, we underline that, while the occurrence of extra mixing during the RGB phase is ascertained firmly by comparisons with observational data for red giant stars, any possible (or probable) signature of these extra mixing phenomena occurring in first-ascent red giants, can be subsequently erased, or masked by similar or even larger isotopic shifts introduced during their AGB stages. Thus, although extra mixing on the RGB was shown to be sufficient to explain the isotopic compositions of *Group* 1 O-rich grains with $^{26}Al/^{27}Al \lessapprox 5\times10^{-3}$ (Palmerini et al. 2017, see their Fig. 2), we cannot exclude the possibility that the extra mixing also extended (caused by the same or a different mechanism) to the AGB phase in the parent stars of these O-rich grains as the majority of dust is expected to form during the AGB phase (see Section 4.2.1.1). Hereafter, we call these extra-mixing processes deep mixing, regardless of the occurrence time (RGB versus AGB) and underlying mechanism (e.g., thermohaline versus magnetic buoyancy). Specifically, deep mixing in low-mass RGB/AGB stars refers to the process during which envelope material is brought down to the stellar interior to allow proton-capture reactions to occur at enhanced stellar temperatures, and, subsequently, the processed material is returned to the envelope.

The higher-than-solar $^{18}O/^{16}O$ isotopic shifts observed in *Group* 1 grains likely reflect initial composition variations. All the FRUITY models (without considering deep mixing) shown in Fig. 6a adopted the terrestrial $^{18}O/^{16}O$ ratio as the initial input and predict a slight destruction of $^{18}O$ during the RGB/AGB phase. If deep mixing occurred in the parent stars of *Group* 1 grains, it could have resulted in further destruction of $^{18}O$ via $^{18}O(p,\alpha)^{15}N$ (Fig. 7a). Thus, the higher-than-predicted $^{18}O/^{16}O$ ratios observed in several *Group* 1 grains (Figs. 5a, 6a, 7) can only be matched if the FRUITY models adopt higher-than-terrestrial $^{18}O/^{16}O$ ratios as the initial input. This, in turn, could be understood if *Group* 1 grains with such $^{18}O/^{16}O$ signatures were derived from RGB/AGB stars of higher-than-solar initial metallicities (Fig. 5a). This inference would be consistent with the previous view that many presolar SiC grains and silicates came from higher-than-solar-metallicity RGB/AGB stars based on their higher-than-solar $^{29}Si/^{28}Si$ and $^{30}Si/^{28}Si$ ratios (see Zinner 2014 and Nittler & Ciesla 2016 for reviews). That the majority of presolar grains came from higher-than-solar-metallicity stars may appear odd, as the grains' parent stars died prior to the solar system formation and are therefore expected to have lower-than-solar metallicities given the general age-metallicity relation (AMR) for nearby stars on the Galactic disc (Delgado Mena et al. 2017; Anders et al. 2018). This conundrum implies that the solar system could have started with a lower-than-average metallicity for its age and/or that dust grains were preferentially made in metal-rich stars, which were more conducive to dust condensation because of their higher Mg, Al, Si contents (*e.g.*, Cristallo et al. 2020). Clearly, further studies are needed to explain why the $^{18}O/^{16}O$ ratio in *Group* 1 oxide grains varies so much.

*4.2.1.3. Mg isotopes*

The close-to-solar Mg isotopic compositions of our *Group* 1 spinel grains are consistent with an origin in RGB/AGB stars of up to 2.0 $M_\odot$, as inferred from their $^{17}O/^{16}O$ ratios (Fig. 6). FRUITY AGB models (Cristallo et al. 2009, 2011) predict <10‰ variations in the Mg isotopic composition during the O-rich phase for low-mass stars (Fig. 6b). In comparison, our 19 *Group* 1 spinel grains show an average composition of $\delta^{25}$Mg = 25 ± 20‰ (1σ Poisson error) and $\delta^{26}$Mg = 13 ± 19‰. However, five of the 19 grains clearly had $\delta^{25}$Mg anomalies (≥ ±100‰ and >2σ errors), while only one of the grains, A15-00-#019, had a significant $\delta^{26}$Mg anomaly, 355 ± 69‰. For grains that fall significantly below the GCE trend line (1:1 line in the 3-Mg isotope plot) expected for the Mg isotopes in Fig. 6b, their large $^{26}$Mg excesses could be resulting from $^{26}$Al decay. Given that our *Group* 1 spinel grain A20-03-#030 (with a large $^{25}$Mg depletion) shows similarities to the silicate grain MET_01B_53_1 of Hoppe et al. (2021) in the O and Mg isotopic compositions, this grain may not have come from low-mass stars.

The more variable $\delta^{25}$Mg values (compared to $\delta^{26}$Mg) observed in our *Group* 1 spinel grains point to the effect of true stellar nucleosynthesis and/or GCE. The fact that most of our *Group* 1 grains lie on the left side of the GCE line (1:1 line) in the 3-Mg isotope plot (though mostly within 2σ errors), may indicate higher-than-predicted $^{25}$Mg production in AGB stars. However, the effect of potential inhomogeneous GCE on Mg isotopes (Nittler 2005) precludes us from drawing any robust conclusion for the following reason. While homogenous GCE models predict that the composition of the ISM varies smoothly because of stellar nucleosynthesis and recycling of stellar materials to the ISM, i.e., yielding an AMR, inhomogeneous GCE models predict that the composition of the ISM varies with time with large scatters and the AMR is disturbed as a result of e.g., heterogeneous mixing of fresh supernova ejecta.

Since the $^{18}O/^{16}O$ ratio and the $\delta^{25}$Mg value are barely altered in low-mass stars according to standard AGB nucleosynthesis models (Cristallo et al. 2009, 2011; Karakas & Lugaro 2016), it was suggested that the initial $^{18}O/^{16}O$ ratios and the $\delta^{25}$Mg values of *Group* 1 grains should be positively correlated as predicted for simple homogeneous GCE models (Nittler et al. 2008). The literature and our spinel grain data, however, do not reveal such a simple correlation (Fig.

8), possibly due to the following complications. (*i*) As will be discussed in Section 4.2.1.3, the parent stars of *Group* 1 grains likely experienced deep mixing that could have destroyed $^{18}$O to certain degrees while enhancing the star's $^{26}$Al production (e.g., Palmerini et al. 2017). (*ii*) Inhomogeneous GCE involving contribution of ejecta/winds from local stars could have affected the initial composition of the grains' parent stars in a manner that would not abide to simple GCE expectation as explained above. The latter is supported by the larger than expected O-Mg isotope heterogeneities observed in *Group* 1 silicate grains by Hoppe et al. (2021) (green open squares in Fig. 8b). The largely variable $^{25}$Mg abundances (up to thousands of ‰ in the δ notation) observed among *Group* 1 silicate grains point to the effects of stellar nucleosynthesis, implying that *Group* 1 silicate grains came from diverse stellar sources (Hoppe et al. 2021). As a result, it is not a simplistic task, if not impossible, to quantify the effects of inhomogeneous GCE on the O and Mg isotopes based on *Group* 1 O-rich grain data.

While *Group* 1 presolar silicates (200–500 nm in size) with high $^{17}$O/$^{16}$O ratios ($\geq 1 \times 10^{-3}$) were found previously to often exhibit large $^{25}$Mg excesses (Fig. 8b), such large $^{25}$Mg excesses were not observed in any of our small *Group* 1 spinel grains, likely reflecting their different stellar origins. Such disparate origins are indeed expected. For example, presolar SiC grains consist of up to ~5-7% supernova grains (AB and X grains; Liu et al. 2017a; Hoppe et al. 2019), while 1/3 of presolar graphite grains likely originated from CCSNe (Amari et al. 2014). The more significant production of graphite relative to SiC by CCSNe has been confirmed by astronomical observations, in the form of abundant amorphous carbon dust but undetectable amount of SiC in supernova remnants (e.g., Gall et al. 2014). Similarly, different types of stars could produce spinel and silicate grains with different spinel-to-silicate ratios, resulting from their slightly different condensation temperatures (1387 K for spinel versus 1346 K for forsterite for gas of solar composition; Lodders 2003) and/or the different elemental compositions required for their formation. Lower condensation temperatures are expected for amorphous silicates, which dominate the presolar silicate population (Nguyen et al. 2007, 2016). The different stellar formation environments required for the formation of spinel and silicate grains are supported by the general lack of *Group* 3 silicate grains (Floss & Haenecour 2016; Nittler et al. 2020), implying that their parent stars produce more oxide than silicate dust compared to other stellar sources. It has been suggested that *Group* 1 presolar silicates with large $^{25}$Mg excesses originated in CCSNe that experienced explosive H-burning in the He/C zone (Leitner & Hoppe 2019) and/or in super AGB stars (Verdier-Paoletti et al. 2019). In either case, the unique stellar environment could have resulted in a preferential formation of silicate over spinel grains as compared to other types of stars, *e.g.*, low-mass RGB/AGB stars; as a result, the rareness of spinel grains from such stars would have precluded us from finding them in this study.

*4.2.1.4. Inferred initial $^{26}Al/^{27}Al$ ratios and deep mixing in AGB stars*

In previous studies of presolar spinel grains (Zinner et al. 2005; Nittler et al. 2008; Gyngard et al. 2010), the initial $^{26}Al/^{27}Al$ ratios of presolar spinel grains were derived based on the $^{26}Mg$ excesses remaining (*i*) after subtracting the contribution of AGB nucleosynthesis along a slope-1.95 line through the origin in the Mg 3-isotope plot for grains with $^{25}Mg$ excesses or (*ii*) after subtracting the contribution of GCE along the slope-1 line for grains with $^{25}Mg$ deficits. The slope-1.95 line adopted by Zinner et al. (2005) was based on FRANEC stellar models. FRUITY stellar models represent an improvement over FRANEC models because the adopted parameters are calibrated against various stellar observations (Cristallo et al. 2009, 2011). Thus, given that FRUITY models for low-mass, close-to-solar metallicity stars predict a negligible effect of AGB nucleosynthesis on the Mg isotopic composition during the O-rich phase (Fig. 6b), we did not correct for the effect of AGB nucleosynthesis and adopted method (*ii*) in this study to correct for the GCE effect to infer the initial $^{26}Al/^{27}Al$ ratio for all our presolar oxide grains. The linear GCE trend with a slope of one for Mg isotopes is generally supported by the Mg isotopic data of *Group* 1 grains recently reported by Hoppe et al. (2021).

Our inferred $^{26}Al/^{27}Al$ ratios are uncertain and possibly inaccurate given the small $^{26}Mg$ excesses observed in most of our spinel grains. With this caveat in mind, we investigate below whether there is any observable difference in the initial $^{26}Al/^{27}Al$ ratio between our small spinel grains and literature oxide grains with more accurately inferred initial ratios based on their larger $^{26}Mg$ excesses because of their higher Al/Mg ratios. By subtracting the expected GCE effect along the slope-1 line, we were able to derive the initial $^{26}Al/^{27}Al$ ratio for 12 of the 25 grains and the 1σ upper limit for another seven grains (Table 1) using the following equation,

$$\frac{26Al}{27Al}=\frac{26Mg_{tot}-24Mg_{tot}\times\left(\frac{26Mg}{24Mg}\right)_{std}-\left[25Mg_{tot}-24Mg_{tot}\times\left(\frac{25Mg}{24Mg}\right)_{std}\right]\times\left(\frac{26Mg}{25Mg}\right)_{std}}{27Al_{tot}\times RSF}$$

in which $^{i}Mg_{tot}$ denotes the total count of isotope $^{i}Mg$ collected in a grain, and the Mg/Al RSF is 0.83 according to the NIST SRM 610 measurements. Thus, $^{26}Mg_{tot}-^{24}Mg_{tot}\times(^{26}Mg/^{24}Mg)_{std}$ and $^{25}Mg_{tot}-^{24}Mg_{tot}\times(^{25}Mg/^{24}Mg)_{std}$ correspond to the $^{26}Mg$ and $^{25}Mg$ excesses, respectively, with respect to their terrestrial abundances, and the numerator of the equation represents the $^{26}Mg$ excess after GCE correction along the 1:1 line in the 3 Mg-isotope plot (in delta notation); the factor of $(^{26}Mg/^{25}Mg)_{std}$ in the last term of the numerator is to correct for the slightly different terrestrial abundances of $^{26}$Mg (11%) and $^{25}$Mg (10%). The errors reported in Table 1 are 1σ Poisson errors.

Our inferred initial $^{26}$Al/$^{27}$Al ratios for the 16 *Group* 1 grains agree well with the literature data as shown in Figs. 7c,d. We chose the isotopic data for *Group* 1 corundum and hibonite grains with large $^{26}$Mg excesses (>100‰ in $\delta^{26}$Mg) for comparison (Choi et al. 1998, 1999; Nittler et al. 1994, 2008), because, given their high Al/Mg ratios, $^{26}$Al decay likely contributed significantly to their $^{26}$Mg excesses. Our *Group* 1 spinel grains overlap well with these literature oxides in the O isotopic ratios and the inferred initial $^{26}$Al/$^{27}$Al ratio (Figs. 7c,d). In comparison to the grain data, the set of FRUITY models shown in Fig. 7c predict the $^{26}$Al/$^{27}$Al production ratio to lie below $4 \times 10^{-3}$ for low-mass stars during the O-rich phase, a factor of three lower than the maximum value inferred from both our and the literature grain data. This discrepancy cannot be explained by uncertainties in relevant nuclear reaction rates because FRUITY model predictions for C-rich low-mass AGB stars, which are the progenitors of presolar mainstream SiC grains (Liu et al. 2018b; Cristallo et al. 2020), yield initial $^{26}$Al/$^{27}$Al ratios that are higher than the inferred ratios of mainstream SiC (mainly between $1\text{–}2\times10^{-3}$; Liu et al. 2021). In other words, the $^{26}$Al-rich *Group* 1 oxide data ($\gtrsim 2\times10^{-3}$) point to higher-than-predicted $^{26}$Al production in their parent stars, while the SiC data suggest lower-than-predicted $^{26}$Al production. Based on FRUITY AGB stellar models, however, we expect lower or comparable $^{26}$Al/$^{27}$Al ratios in *Group* 1 O-rich grains compared to MS SiC grains because the $^{26}$Al/$^{27}$Al ratio at the surface is predicted to increase during the stellar evolution from O-rich to C-rich phase (Fig. 7c), and is higher in $\gtrsim 1.5\ M_\odot$ low-mass stars (parent stars of O-rich and C-rich grains) than in $\lesssim 1.5\ M_\odot$ (parent stars of O-rich grains) because of increasing H-burning temperature with increasing initial stellar mass.

The data-model discrepancy for Group 1 O-rich grains and mainstream SiC grains implies that the parent RGB/AGB stars of *Group* 1 grains with high initial $^{26}$Al/$^{27}$Al ratios likely experienced deep mixing, resulting in enhanced $^{26}$Al productions at high temperatures in their parent stars. Based on observed low $^{12}$C/$^{13}$C ratios in low-mass RGB stars, deep mixing processes were proposed to occur in low-mass RGB stars leading to enhanced production of $^{13}$C via $^{12}C(p,\gamma)^{13}N(\beta^{+}\nu)^{13}C$ (e.g, Wasserburg et al. 1995). Deep mixing was first proposed to occur during the RGB phase based on a parameterized mixing model, reaching down to $(3\text{–}3.5)\times10^{7}$ K (Wasserburg et al. 1995). Later on, physical mechanisms (e.g., thermohaline diffusion) that could power such deep mixing during the RGB phase were proposed (Eggleton et al. 2006; Charbonnel and Zahn 2007). However, temperatures $\gtrsim 5\times10^7$ K are required to produce $^{26}Al/^{27}Al>5\times10^{-3}$ (Nollett et al. 2003; Palmerini et al. 2011). Thus, deep mixing processes during the AGB phase are favored over those during the RGB phase to explain the isotopic compositions of *Group* 1 grains with $^{26}Al/^{27}Al > 5\times10^{-3}$.

In parameterized deep-mixing models, the proton-capture reactions during deep mixing are controlled by two factors: the mass circulation flow rate and the maximum temperature ($T_{max}$) reached at the bottom of the radiative zone in the stellar interior. Previous deep-mixing calculations (Nollett et al. 2003; Palmerini et al. 2011) showed that (*i*) the final $^{26}Al/^{27}Al$ ratio in the stellar envelope depends solely on $T_{max}$ experienced by the circulating material during deep mixing, when $^{26}Al$ is produced via $^{25}Mg(p,\gamma)^{26}Al$ and (*ii*) the final $^{18}O/^{16}O$ ratio in the stellar envelope is mainly sensitive to the circulation rate during the deep-mixing process, when $^{18}O$ is destroyed via $^{18}O(p,\alpha)^{15}N$. Different from these previous, parametrized deep-mixing models, the deep-mixing model of Palmerini et al. (2017, 2021) is a physical model that considers the effect of magnetic buoyancy on deep mixing in RGB/AGB stars. In this model (Fig. 7), deep mixing of envelope material is powered by advection of magnetic bubbles into the envelope, which differs from the classical forms of deep mixing based on thermohaline diffusion. In the classic view of deep mixing induced by thermohaline diffusion, a conveyor belt circulation is triggered by a downward flow of material from the border of the convective envelope into the radiative region, thus driving an upward flow of matter. In the magnetic-buoyancy-induced-mixing scenario, the diffusive downflow of material is triggered by the relatively fast upflow of the magnetic bubbles, and the mixing velocity is determined by the velocity at which magnetized bubbles cross the region between the H-burning shell and the base of the convective envelope, $v(r) = v(r_k)\left(\frac{r_k}{r}\right)^{k+1}$, in which $r$ is the position along the stellar radius and $k$ is the index of the power law $\rho \propto r^k$, which is related to the density distribution in the crossed region. As explained by Palmerini et al. (2017), $k$ can be used to determine the deepest layer from which the mixing starts, and $r_k$ and $v(r_k)$ denote the mixing starting depth and velocity, respectively. The smaller the $k$ value, the deeper the mixing, the larger the $T_{max}$ value, and the larger the circulation rate. Thus, in Fig. 7 we see that with decreasing $k$ value, the model of Palmerini et al. (2021) predicts increasing $^{26}Al/^{27}Al$ (resulting from increasing $T_{max}$) and decreasing $^{18}O/^{16}O$ (resulting from increasing circulation rate) ratios.

The deep-mixing models of Palmerini et al. (2021), however, cannot consistently explain the O and Al isotopic compositions of the $^{26}$ Al-rich *Group* 1 grains (Figs. 7a,c). Nollett et al. (2003) showed that when the circulation rate lies below $10^{-7}$ $M_{\odot}$/year, the $^{18}O/^{16}O$ ratio remains almost unchanged in the stellar envelope with increasing $T_{max}$ because very little material is processed over the lifetime of the AGB for these mixing rates. Thus, for the model of Palmerini et al. (2021) to explain the $^{26}$Al-rich *Group* 1 grain data, it requires a significant reduction in the circulation rate without significantly affecting the $T_{max}$ values achieved in the $k = -3.3$ and $k = -3.5$ models. This requirement seems unachievable given the underlying relationship between the circulation rate and $T_{max}$ resulting from magnetic buoyancy. This needs further investigation however, which would involve running more tests for AGB stars with a wide range of initial masses and metallicities. In conclusion, the comparable $^{18}O/^{16}O$ ratios between $^{26}$Al-rich ($\gtrsim 5\times10^{-3}$) and $^{26}$Al-poor *Group* 1 grains (Figs. 7c,d) suggest that the $^{26}$Al-rich *Group* 1 grains formed in AGB stars that experienced deep mixing at high $T_{max}$ ($\gtrsim 5\times10^{7}$ K) but extremely low circulation rates ($<10^{-7}$ $M_{\odot}$/year).

*4.2.2. Group 2 spinel grains: deep mixing versus HBB*

Compared to *Group* 1 grains, the larger $^{18}$O depletions of *Group* 2 grains are believed to result from (*i*) the operation of deep mixing at higher circulation rates in low-mass RGB/AGB stars (Wasserburg et al. 1995; Nollett et al. 2003; Palmerini et al. 2011) and/or (*ii*) proton-capture at the base of the convective envelope at sufficiently high temperatures, i.e., HBB, in intermediate-mass AGB stars (Lugaro et al. 2017). Compared to deep mixing, HBB can generally reach higher temperatures for proton-capture reactions to occur, thus leading to significant $^{25}$Mg production via $^{24}Mg(p,\gamma)^{25}Al(\beta^{+}\nu)^{25}Mg$ above $1 \times 10^{8}$ K (the upper limit of what can be achieved by deep mixing).

Figures 7a,b shows that our four *Group* 2 grains could have originated in 1.5 $M_{\odot}$ low-mass AGB stars (Fig. 7a) or 4.5–6 $M_{\odot}$ intermediate-mass stars. However, the low inferred initial $^{26}Al/^{27}Al$ ratios ($<5\times10^{-3}$) of our *Group* 2 grains favor a low-mass stellar origin, 1.5 $M_{\odot}$. This is because while the O isotopic compositions of the four *Group* 2 grains can be explained by mixing 60–70% final HBB stellar envelope material with 40–30% solar material (Fig. 7b), the same mixtures are predicted to have $^{26}Al/^{27}Al$ ratios above $1\times10^{-2}$ (Fig. 7d), too high to explain our grain data. The predicted high initial $^{26}Al/^{27}Al$ ratios could reflect modelling uncertainties, (e.g., too high $T_{max}$ for HBB), but we found that the HBB models of Lugaro et al. (2017) encounter the same problem in explaining our Group 2 grain data. In their model, the O isotopic

data require mixing 70−80% of the 6 $M_\odot$ HBB products at TP #22−34 with 30−20% solar system material, but all these mixtures are predicted to exhibit $>1\times10^{-2}$ $^{26}Al/^{27}Al$ ratios, consistent with the conclusion based on our HBB models. Also, the mixtures required by the O isotopic data based on the two sets of HBB models both predict a wide range of $\delta^{25}Mg$ values: ~100−1500‰ by our HBB models and ~100−1300‰ by those of Lugaro et al. (2017). Although all of our four *Group* 2 grains lie above the solar $^{25}Mg/^{24}Mg$ ratio in the 3-Mg isotope plot, their average $\delta^{25}Mg$ value is only 31 ± 25‰ (1σ Poisson error), which agree roughly with the predicted lower limit (~100‰).

In comparison, *Group* 2 grains with high $^{17}O/^{16}O$ ($\gtrsim1\times10^{-3}$), low initial $^{26}Al/^{27}Al$ ($<1\times10^{-2}$) and close-to-normal $\delta^{25}Mg$ ratios are better explained by deep mixing occurring in low-mass RGB/AGB stars (Figs. 7a,b). In the deep-mixing scenario (Fig. 7c), the different $^{26}Al/^{27}Al$ ratios of *Group* 2 grains can be explained by different maximum temperatures reached by the circulating materials during the deep mixing in their parent stars: the smaller the $k$ value, the higher the maximum temperature, and the higher the $^{26}Al/^{27}Al$ ratio. The average $\delta^{25}Mg$ value of our four grains, 31 ± 25‰ (1σ Poisson error), is also consistent with the <10‰ $\delta^{25}Mg$ variations by RGB/AGB nucleosynthesis in low-mass stars predicted by the FRUITY models (Fig. 6b). The $\delta^{25}Mg$ value at the stellar surface cannot be modified by deep-mixing processes due to the low $T_{max}$ that can be achieved in low-mass stars ($< 1 \times 10^8$ K).

Finally, we see in Fig. 7d that some of the *Group* 2 grains from the literature indeed had initial $^{26}Al/^{27}Al$ ratios above 0.01 and could have originated from intermediate-mass AGB stars (Lugaro et al. 2017). To definitively confirm the stellar site that produced *Group* 2 grains, isotopic compositions of *s*-process elements measured with Resonance Ionization Mass Spectrometry (RIMS) (Stephan et al. 2016) can provide more insights. This is because (*i*) the production of isotopes of heavy elements sitting close to branching points along the *s*-process, e.g., $^{96}Zr$, $^{134,137}Ba$, is controlled by the neutron flux produced by the minor neutron source for the *s*-process, $^{22}Ne(\alpha, n)^{25}Mg$ reaction, and is insensitive to the initial stellar composition and (*ii*) the activation of the $^{22}Ne(\alpha, n)^{25}Mg$ reaction during TPs is controlled by the maximum temperature, which increases with increasing initial stellar mass (Lugaro et al. 2003; Liu et al. 2014a, 2014b). Thus, the signatures of isotopic ratios that are affected by *s*-process branching points in *Group* 2 grains can be used to better distinguish between low- and intermediate-mass stellar origins (*e.g.*, Liu et al. 2019).

*4.2.3. Group 3 grains*

*Group* 3 oxide grains are thought to have originated in low-mass, low-metallicity AGB stars based on the assumption that their lower-than-solar $^{17}O/^{16}O$ and $^{18}O/^{16}O$ ratios reflect the GCE trend for O isotopes (see Nittler & Ciesla 2016 for a review). Compared to the slope-1 GCE trend in the O 3-isotope plot (in delta notation), *Group* 3 oxide grains generally show higher $^{17}O/^{16}O$ ratios, likely pointing to the production of $^{17}O$ in the grains' parent stars as discussed in Section 4.2.1. However, very few *Group* 3 grains have been analyzed for isotopic compositions of elements other than O, and their stellar origins await confirmation. Unfortunately, we did not find any good *Group* 3 candidates from this study for further Al-Mg isotope analysis.

### *4.3. Presolar Spinel Grains from Novae*

Novae are powered by thermonuclear explosions in the H-rich envelopes of WDs. These H-rich envelopes are produced by transfer of material from a low-mass stellar companion that is still on the main sequence or on the giant branch. There are two types of classical novae, depending on the nature of the WD: CO WDs are the remnants of evolved AGB stars less massive than ~6–8 $M_\odot$, and ONe WDs are remnants of more massive AGB stars (8–10 $M_\odot$) (José et al. 2004; Herwig 2005; Karakas & Lattanzio 2014). Nova nucleosynthesis models predict ubiquitous production of $^{13}C$, $^{15}N$, $^{17}O$, $^{25}Mg$, $^{26}Mg$, and $^{26}Al$ by explosive H burning in both massive CO (M ≥ 0.8 $M_\odot$) and ONe novae (José and Hernanz 2007). Since RGB/AGB stars are predicted to produce $^{17}O/^{16}O$ ratios up to ~0.005 (Fig. 6a), grains with larger $^{17}O$ excesses cannot be explained by AGB nucleosynthesis but are in line with the signature of explosive H burning at high temperatures (> $1 \times 10^8$ K) in novae (Nittler et al. 2008; Gyngard et al. 2010). Thus, grains with $^{17}O/^{16}O$ ratios above 0.005 are considered as putative nova grains.

The isotopic data of three spinel grains from this study and Gyngard et al. (2010) and two silicate grains from Nguyen & Messenger (2014) are compared to the mean (averaged by mass) nova compositions predicted by the CO and ONe nova models of José & Hernanz (2007) in Fig. 9. Their O isotopic signatures point to similar parent novae (Fig. 9a). We did not include the nova grain from Leitner et al. (2012), as its O isotopic signature would point to a parent CO nova of lower initial mass (0.6 $M_\odot$< M < 1.0 $M_\odot$). The nova models in Fig. 9 were previously adopted in the studies of Liu et al. (2016, 2017a) and Boujibar et al. (2021) for comparison with nova SiC grains. The data-model comparison for O and Mg isotopes in Fig. 9 reveals that the nova ejecta needs to be largely diluted with ISM materials of solar isotopic composition to match the isotopic composition of the nova grains. Our mixing calculations show that the O isotopic compositions of the nova grains can be explained by mixing 95−50% ISM material with 5%−50% of the ejecta (the mixing ratio is calculated based on $^{16}O$) of a 1.15 $M_{\odot}$ CO nova that has an initial composition with a 25% degree of mixing between the core material and accreted envelope. The diluted 1.15 $M_{\odot}$ CO nova ejecta, however, cannot simultaneously explain the Mg isotopic compositions of any of the five nova grains (Fig. 9): (*i*) our spinel grain requires mixing with 30% nova ejecta for its O isotopic composition while its Mg isotopic composition needs to be explained by mixing with ~10% nova ejecta for an Al/Mg ratio of 1.2 (yielding $\delta^{25}Mg$ = 20‰ and $\delta^{26}Mg$ = 69‰), (*ii*) the spinel grain G10-1 (Gyngard et al. 2010) requires mixing with 50% nova ejecta for its O isotopic ratios but 80% for its Mg isotopic ratios; also the data does not fall onto the trend (brown solid line) that accounts $^{26}Al$ decay in spinel (with Al/Mg =2), (*iii*) the silicate grain N14-2 (Nguyen & Messenger 2014) requires mixing with 20% nova ejecta for its O isotopic composition but 80% mixing for its Mg isotopic composition, (*iv*) the O and Mg isotopic signature of grain N14-1 (Nguyen & Messenger 2014) suggests mixing with 25% and 50% nova ejecta, respectively, (*iv*) the spinel grain G10-2 (Gyngard et al. 2010) requires 5% mixing for its O isotopes, which would yield an ejecta with $\delta^{25}Mg$ = 10‰ and $\delta^{26}Mg$ = 1‰; however, the grain is depleted in $^{25}Mg$, which could be explained if the accreted ISM material or the CO WD started with such an initial Mg isotopic composition. Thus, although the O and Mg isotopic signatures of the nova grains can be qualitatively explained by the CO nova models of José & Hernanz (2007), in four of the five cases (*i*-*iv*) the nova models cannot quantitatively explain the O and Mg isotopic compositions of the nova grains (discrepancies by up to a factor of four). Iliadis et al. (2018) investigated the probabilities of C-rich and O-rich nova grains being sourced from novae by exploring a large range of model parameters (e.g., WD composition, peak temperature and density) based on a Monte Carlo technique. Their study found low probabilities for the O-rich nova grains from Gyngard et al. (2010) and Nguyen & Messenger (2014) to have come from novae. Thus, it remains a question whether the data-model inconsistencies reflect uncertainties in involved nuclear reaction rates or these extremely $^{17}O$-rich grains did not come from novae.

### *4.4. Presolar Spinel Grains from CCSNe*

CCSN is an explosive event during which the Fe core of a massive star collapses and then rebounds, which results in shock heating the outer layers of the star and the explosion of the star outward. A massive star with the initial stellar mass above ~10 $M_{\odot}$ undergoes this type of explosion (Woosley & Weaver 1995). *Group* 4 grains are inferred to have originated from CCSNe based on their O isotopic signatures. The trend defined by our *Group* 4 oxides (Fig. 1a) can be explained by mixing material from the inner $^{16}$O-rich zones with material from the outer He/C zone and the envelope (Nittler et al. 2020). The $^{16}$O-rich signature of the inner zones results from a series of alpha capture, *e.g.*, $^{12}$C($\alpha$, $\gamma$)$^{16}$O during He burning, and the reversed reactions, *e.g.*, $^{20}$Ne($\gamma$, $\alpha$)$^{16}$O during Ne burning, during the pre-supernova phase. The outer He/C zone is enriched in $^{18}$O by partial He burning in the He/C zone via $^{14}$N($\alpha$,$\gamma$)$^{18}$F($\beta^+ \nu$)$^{18}$O, while the envelope is enriched in $^{17}$O via $^{16}$O($p$, $\gamma$)$^{17}$F($\beta^+ \nu$)$^{17}$O. The CCSN origin of *Group* 4 grains is supported by the multielement isotopic data for two *Group* 4 oxides reported by Nittler et al. (2008) and for eight *Group* 4 silicates by Nguyen & Messenger (2014). In addition, Hoppe et al. (2021) recently proposed that $^{25}$Mg-poor *Group* 1 grains could have been sourced from pre-supernova massive stars and/or CCSNe. Given its large $^{25}$Mg-depletion, our *Group* 1 spinel grain A20-03-#030 therefore may represent dust from a pre-SN massive star or CCSN instead of a low-mass star.

Compared to the larger oxide grains from the literature (Fig. 1), our spinel grain population (~200 nm on average) seems to contain a larger fraction of *Group* 4 grains (about 2σ significance). This observation is in line with the following observations: (*i*) Hoppe et al. (2015) found that *Group* 3 and *Group* 4 silicates from CCSNe are more abundant in the 100–200 nm size range than the larger size range and (*ii*) Hoppe et al. (2010) found that type C SiC grains from CCSNe were more abundant in the 0.2–0.5 μm size range than micrometer-size range. This implies that CCSNe, on average, produce smaller dust grains than solar-metallicity AGB stars, regardless of the condensation environment (O-rich versus C-rich). The enhanced production of small dust grains in SNe could be caused by the higher velocities of gas from the interior supernova zones and the larger temperature and pressure gradients in the mixed supernova ejecta from which the grains condensed. Such a dust condensation environment in the CCSN ejecta was also suggested to explain the unique structural features of SiC X grains from CCSNe compared to SiC grains from other types of stars, e.g., higher concentrations of subgrains/impurity solid solutions and stacking defects and an enhanced percentage of higher-order non-3C SiC polytypes observed in X grains (*e.g.*, Liu et al. 2017b; Singerling et al. 2021).

## 5. Conclusions

Primitive extraterrestrial materials contain presolar grains formed in the outflows of stars that died prior to solar system formation. The stellar origins of these presolar grains are some of the key questions in astrophysics and cosmochemistry. To further our understanding of the presolar inheritance of the solar system, we investigated the O and Al-Mg isotopic systematics of oxide grains that were chemically separated from the CI carbonaceous chondrite. A comparison with state-of-the-art stellar models for low-to-intermediate-mass AGB stars suggests that our *Group* 1 and *Group* 2 grain data are more consistent with originating from low-mass RGB/AGB stars. The parent stars of our *Group* 1 grains with $(^{26}Al/^{27}Al)_0 > 5\times10^{-3}$ and *Group* 2 grains likely experienced deep-mixing processes but under different conditions. Indeed, the lack of large $^{18}O$ depletions in *Group* 1 grains points to the occurrence of deep mixing at much lower circulation rates in their parent stars than in the parent stars of *Group* 2 grains. The different $^{25}Mg$ isotopic signatures between *Group* 1 spinel and silicate grains provide further evidence that *Group* 1 silicate grains with large $^{25}Mg$ excesses (Leitner & Hoppe 2019; Verdier-Paoletti et al. 2019; Hoppe et al. 2021) did not originate in low-mass RGB/AGB stars. The large $^{25}Mg$ depletion identified in an extremely $^{17}O$-rich *Group* 1 grain suggests that this grain could have been sourced from a pre-supernova massive star or CCSN. The O isotopic compositions of the nova grains from this study are consistent with a massive CO nova origin, but the Mg isotopic composition of grain A15-08-#009 cannot be simultaneously explained by the same CO nova-ISM mixture based on the CO nova models of José & Hernanz (2007), which is a common problem when explaining the Mg-Al isotopic systematics of literature nova O-rich grains using existing nova models. Finally, we also observed that our small spinel grains comprise a higher percentage of supernova grains than the larger presolar oxides reported in the literature, implying that the relative supernova grain abundance increases with decreasing grain size, in line with the trends previously observed for *Group* 3 and *Group* 4 silicates and type C SiC grains from supernovae. The enhanced production of small *Group* 4 spinel grains could be caused by the dynamic dust condensation environment in CCSN ejecta.

**Supplementary Materials:** Table A1: Oxygen isotopic compositions of additional presolar oxides from Orgueil acid residue; Figs. A1-A4; Presolar Grain Selection Criterion.

**Acknowledgments:** We thank the Associate Editor Yves Marrocchi and three anonymous referees for their careful and constructive reading of the manuscript. We also would like to thank Dr. Ryan Ogliore for reading an earlier version the manuscript. This research was funded by NASA Emerging Worlds program 80NSSC20K0387 to N. L.

**Conflicts of Interest:** The authors declare no conflict of interest.

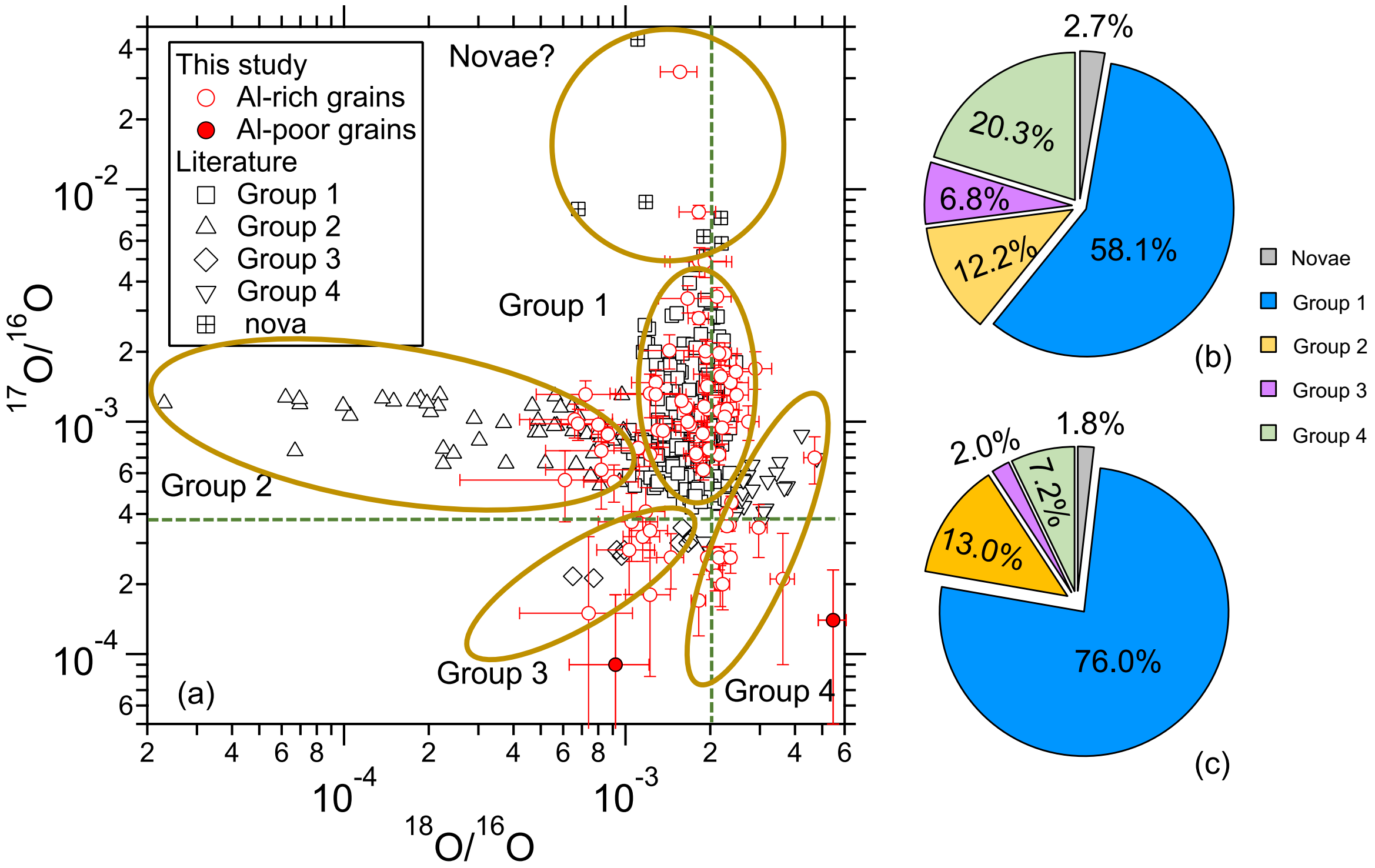

**Figure 1**. (a) Oxygen isotopic ratios of our new presolar O-rich grains from the Orgueil acid residue prepared by Dauphas et al. (2010), compared with literature data for presolar oxides of larger size (Presolar Grain Database; Hynes & Gyngard 2009; see text for details; available as supplementary only material). Ellipses are based on the grain groups redefined by Nittler et al. (2020), and green dashed lines represent terrestrial O isotopic ratios. Panels (b) and (c) illustrate the percentages of the five groups of presolar oxide grains in this study (panel b) and in the literature (panel c). Note that we reclassified all the literature data based on the redefined grain groups given by Nittler et al. (2020). Error bars in this and subsequent figures are all 1 σ.

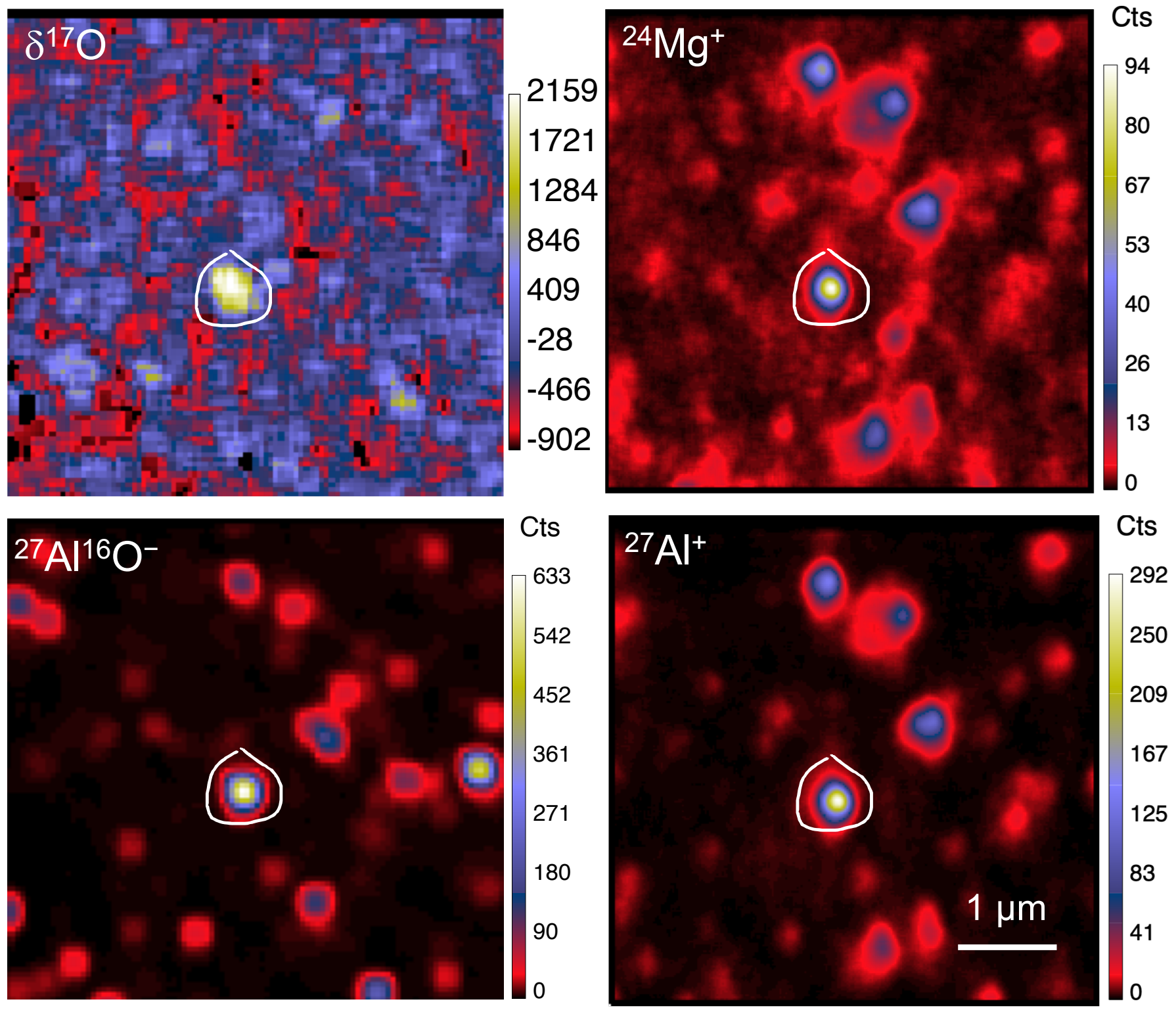

**Figure 2**. Illustration of using NanoSIMS isotope and ion images of *Group* 1 spinel grain A05-07-#011 to relocate the grain for Al-Mg isotopic analysis. The $\delta^{17}$O isotope image was produced by binning five pixels to increase the statistics using L'image. The grain is highlighted by white contour lines in all the images.

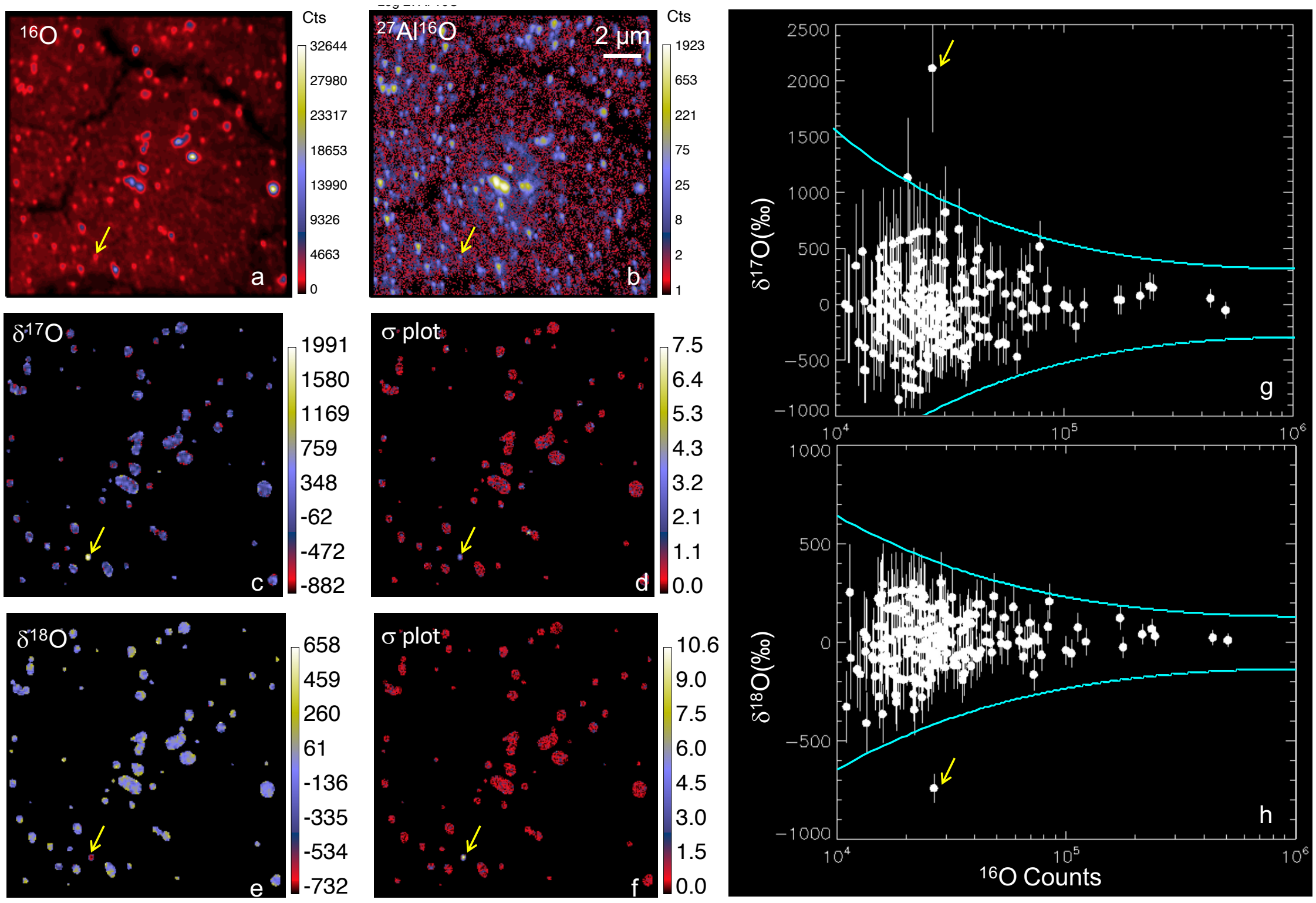

**Figure 3**. Oxygen isotopic data reduction using L'image software. Shown in panels a-b are $^{16}O^{-}$ and $^{27}Al^{16}O^{-}$ secondary ion images, in panels c-d are the $\delta^{17}O$ isotopic image and corresponding σ plot, in panels e-f are the $\delta^{18}O$ isotopic image and corresponding σ plot, and in panels g-h are $\delta^{17}O$ and $\delta^{18}O$ values of 206 O-rich grains that were automatically identified in the $^{16}O$ ion image (panel b) using the L'image. In panels g-h, also plotted are ±3 σ range of data (cyan solid lines) as a function of the total $^{16}O$ count, derived from the data. A *Group* 2 Al-rich grain is highlighted by yellow arrows in all the panels.

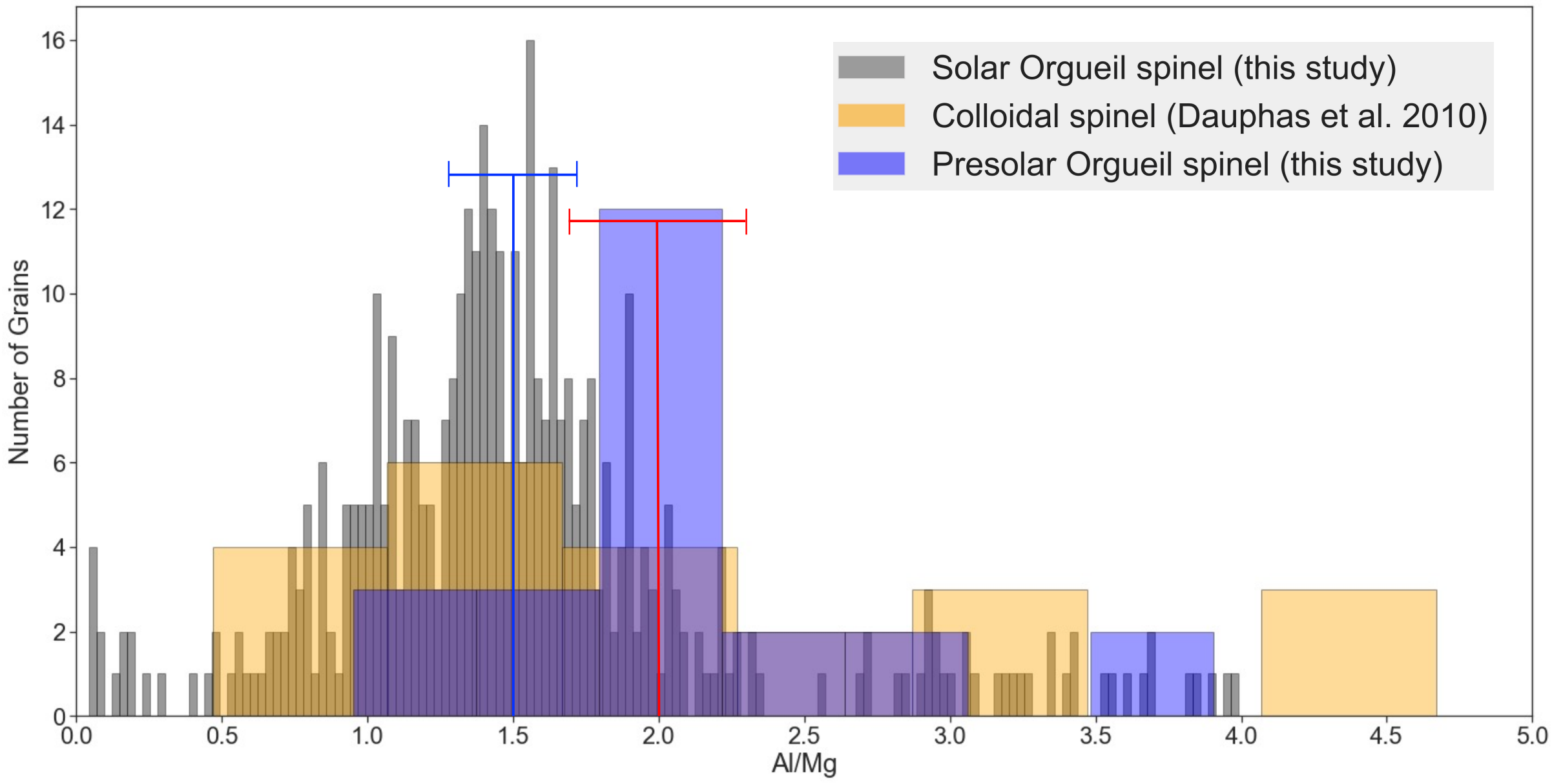

**Figure 4**. A comparison of 24 O-anomalous spinel grains (Table 1) with all other Al-rich grains in the same $Al^+$ ion images for their Al/Mg ratios. Also shown are the TEM-EDX Al/Mg ratios of Al-rich spinel grains from Orgueil and Murchison acid residues reported in Table 1 of Dauphas et al. (2010). The Al/Mg data from this study were corrected for the Mg/Al RSF based on SRM 610 measurements. The red and blue lines represent the 15% 1σ errors in the determined Al/Mg ratio for Al/Mg ratios of 2 and 1.5, respectively.

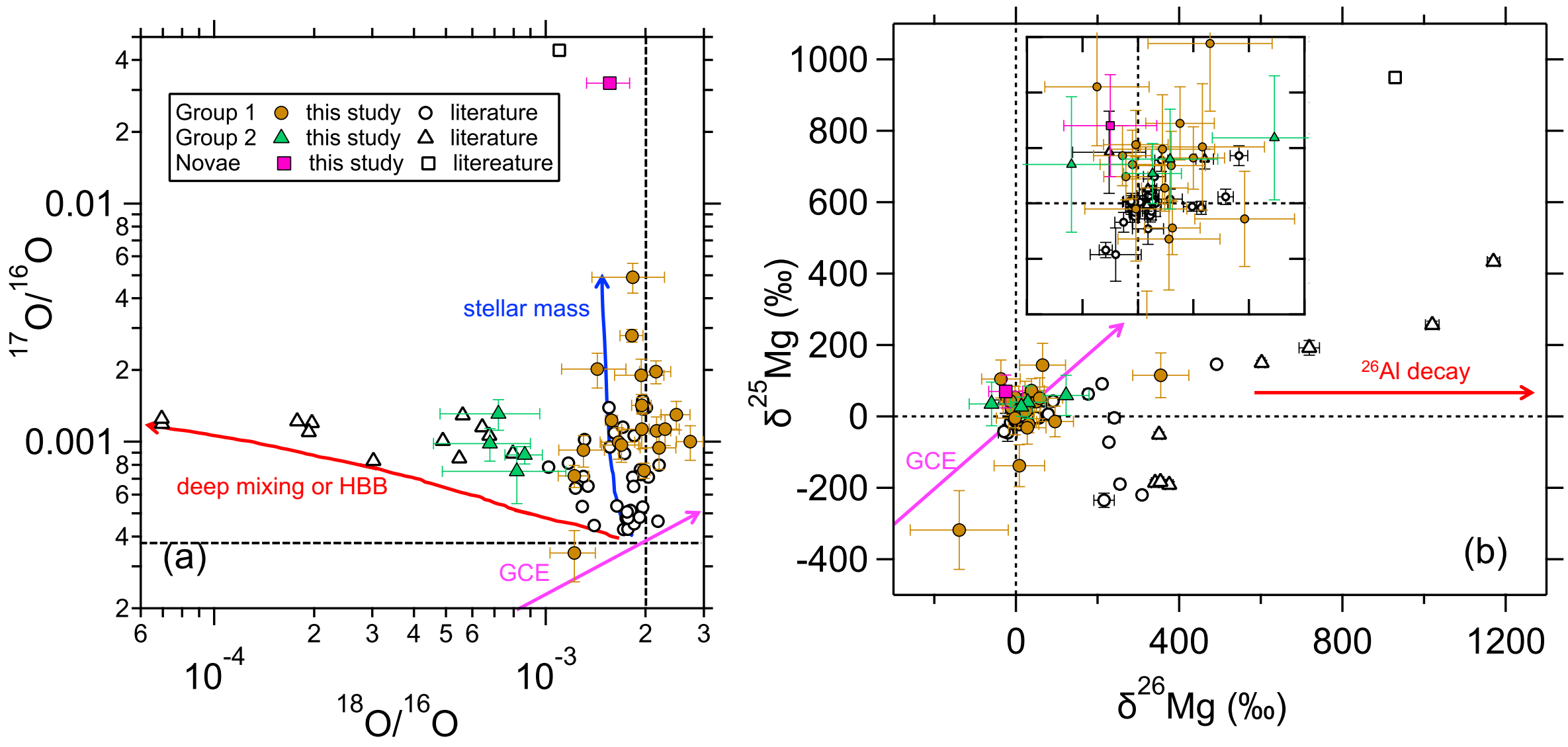

**Figure 5.** A comparison of presolar spinel grains isotopic ratios from this study with literature data (Zinner et al. 2005; Nittler et al. 2008; Gyngard et al. 2010) for their O (panel a) and Mg (panel b) isotopic ratios. The effects of different astrophysical processes (denoted by colored text) are illustrated using colored curves (for illustration purpose, not actual model predictions). HBB stands for hot bottom burning, and GCE Galactic chemical evolution. The dashed lines denote terrestrial isotopic ratios. Panel b includes a subpanel that zooms into the dense grain region (−100‰ to 150‰ in both x and y axes).

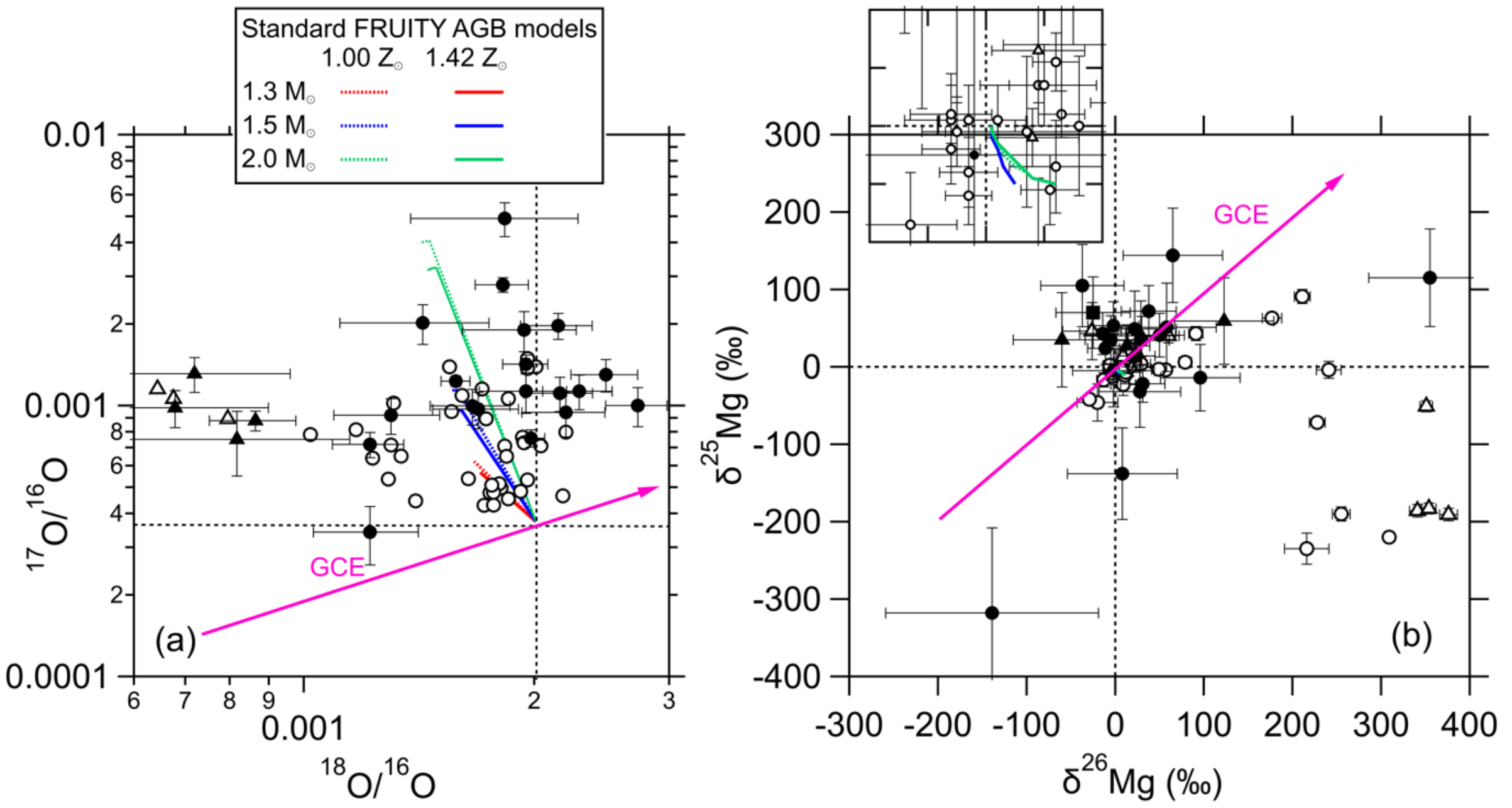

**Figure 6.** A comparison of the same set of presolar spinel grains data as in Fig. 5 with the FRUITY model predictions of Cristallo et al. (2009, 2011) for low-mass RGB/AGB stars during the O-rich phase. Note that (*i*) the x and y axis ranges are narrower in both panels compared to Fig. 5 and (*ii*) different groups of grains from this study are all shown as filled symbols in black instead of in different colors as in Fig. 5. Panel b includes a subpanel that zooms into the FRUITY model region (−20‰ to 20‰ in both x and y axes).

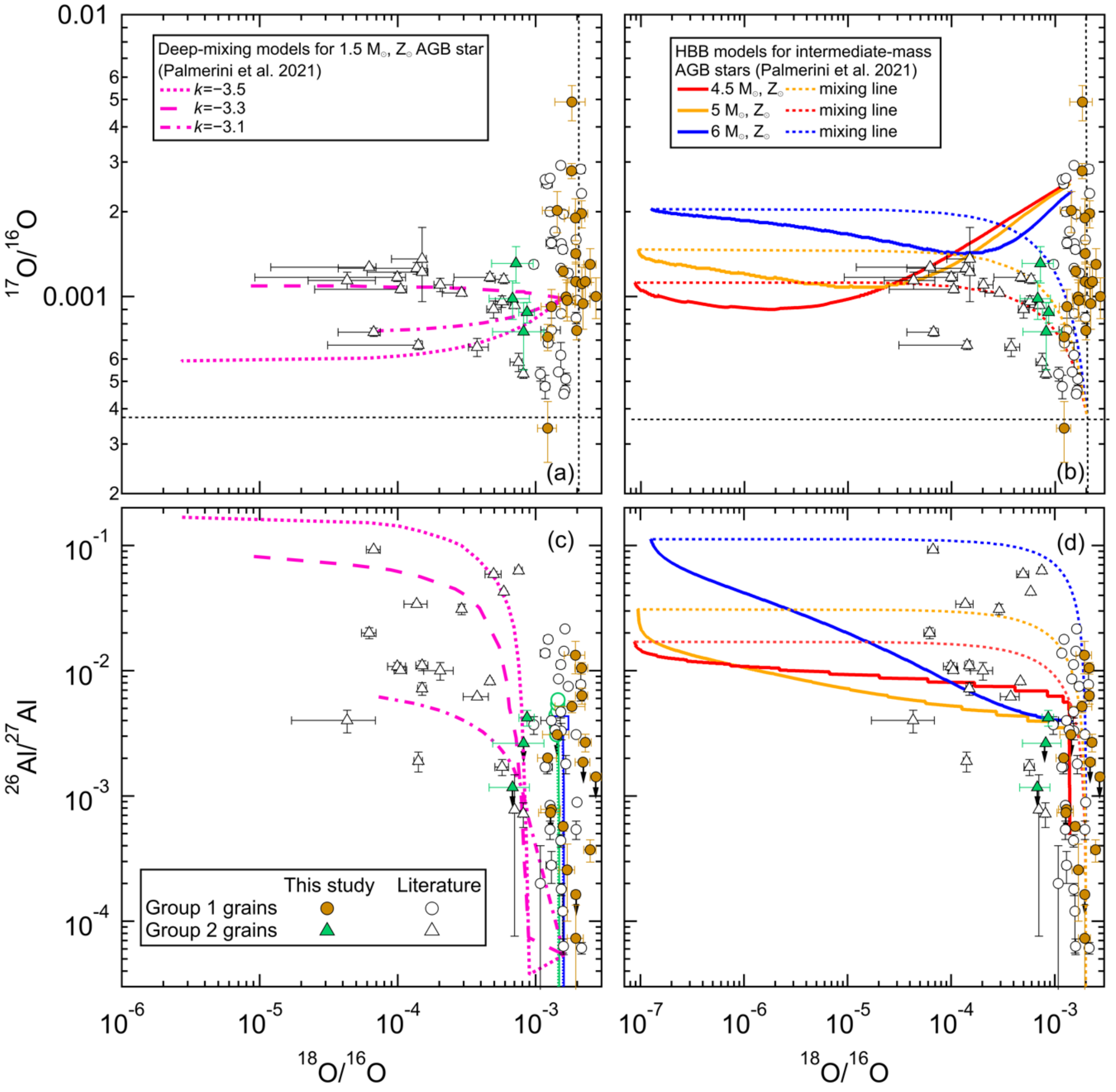

**Figure 7.** Comparison of *Group* 1 and *Group* 2 oxide grain data from the literature (mostly from Choi et al. 1998, 1999; Nittler et al. 1994, 2008) and this study. The data are also compared to predictions for the deep-mixing (panels a and c) and HBB (in panels b and d) models of Palmerini et al. (2021) (see the main text for details). The same set of FRUITY models (the C-rich phase is also included here as symbols) as shown in Fig. 6 is also included in panel c for comparison with the grain data. The initial compositions of the deep-mixing and HBB models are those predicted by the corresponding FRUITY stellar models after the FDU.

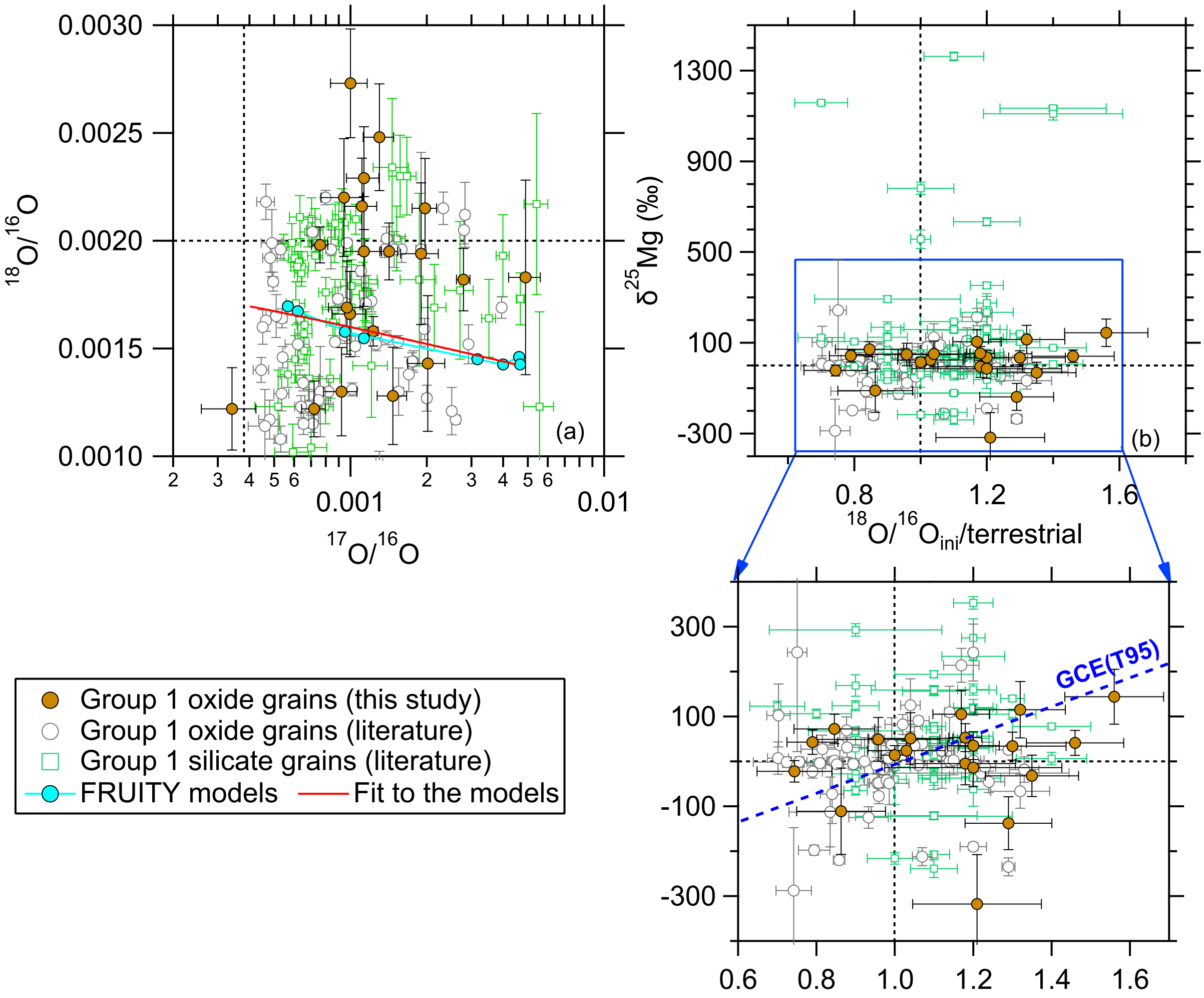

**Figure 8.** Comparison of O and Mg isotopic data in *Group* 1 spinel grains (Zinner et al. 2005; Nittler et al. 2008; Gyngard et al. 2010) and silicate grains (Hoppe et al. 2021) from the literature, and spinel grains from this study. The FRUITY model predictions for the envelope composition in low-mass (1.3–2.5 $M_\odot$), close-to-solar-metallicity (1.0–1.4 $Z_\odot$) RGB/AGB stars during the O-rich phase (cyan line with dots in panel a, fitted (red solid line) using the equation, $\log\left(\frac{18O}{16O}\right)_{fit} = -0.0745 \times \log\left(\frac{17O}{16O}\right)_{fit} - 3.0204$ with $R^2$ =0.94). The initial $^{18}O/^{16}O$ ratio in panel (b) was calculated based on the fit line using the equation $(\frac{18O}{16O})_{initial} = (\frac{18O}{16O})_{grain} + [(\frac{18O}{16O})_{terrestrial} - (\frac{18O}{16O})_{fit}]$. The GCE model predictions by Timmes et al. (1995) (blue dashed line) are shown in the zoom-in panel of panel (b) for comparison.

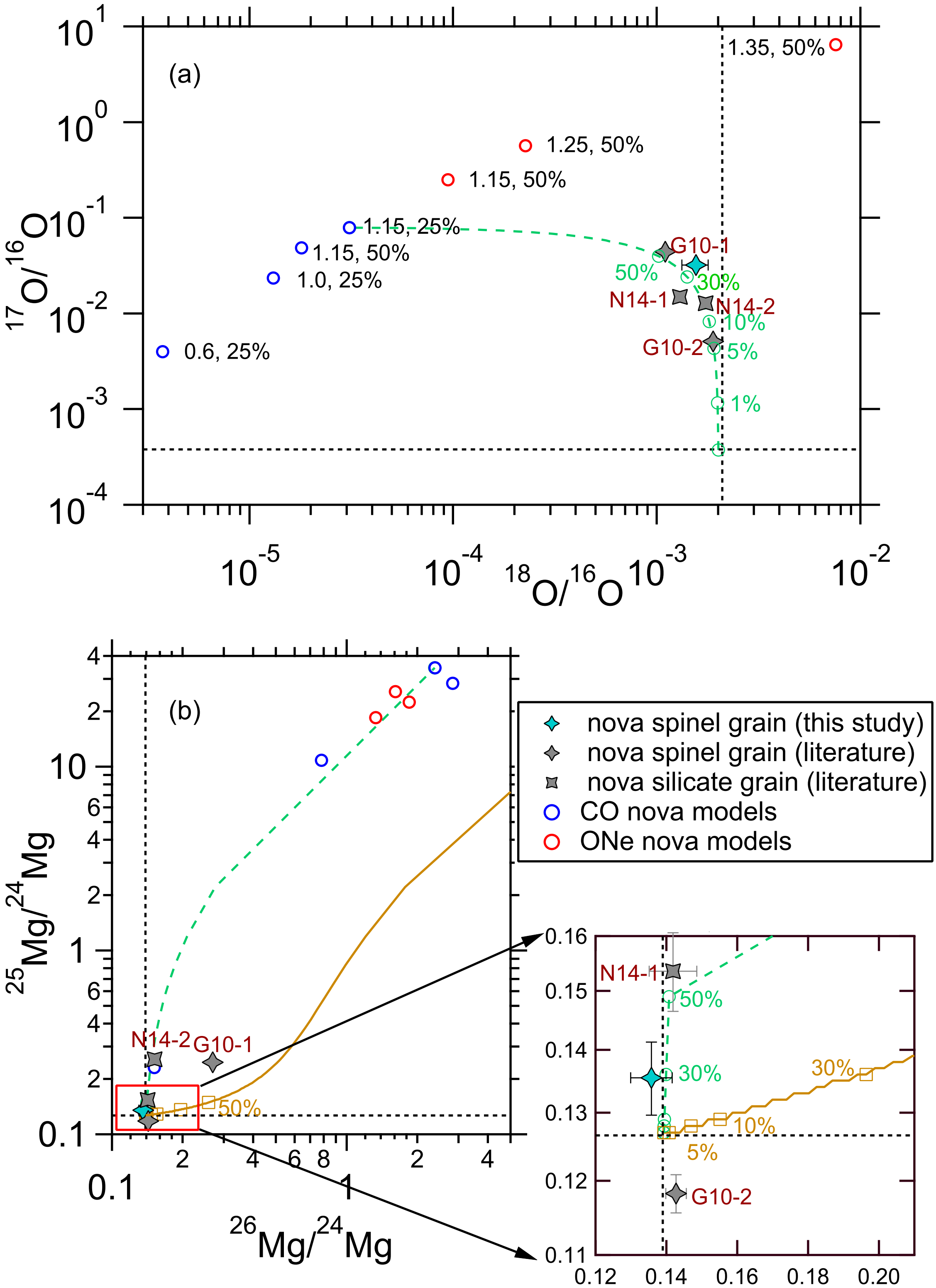

**Figure 9**. Nova model predictions of José & Hernanz (2007) are compared to the nova grain data from this and previous (Gyngard et al. 2010; Nguyen & Messenger 2014) studies. The two nova grains from Gyngard et al. (2020) are labeled with G10-1 and G10-2 and grains from Nguyen & Messenger (2014) N14-1 and N14-2. The terrestrial values are denoted by dashed lines in the plots. The pair of numbers next to each model denote the initial stellar mass of the WD (in solar mass, the first number) and the degree of mixing between the core material and accreted envelope (the second number). Mixing calculations with ISM material (assuming solar composition given in Lodders 2003) are shown for the 1.15 CO nova model (25% degree of mixing) with the number (in percentage) indicating the amount of nova ejecta mixed with the ISM material, from which the grains could have condensed. For $^{26}Mg/^{24}Mg$, the green line denotes the case where the $^{26}Mg$ budget of a grain did not receive any contribution from $^{26}Al$ decay (for comparison with nova silicate grains), and the brown line denotes the case where spinel grains (with Al/Mg =2) condensed from the nova ejecta (for comparison with nova spinel grains); note that the Al/Mg enrichment factor of the spinel dust relative to the ejecta varies from 24 for the solar composition to 1 for the nova composition.

**Table 1.** Isotopic compositions of presolar oxides from Orgueil acid residue.

| Grain Name | Group | Phase | Diameter[1] (nm) | $^{17}O/^{16}O$ ($\times10^{-4}$) | $^{18}O/^{16}O$ ($\times10^{-4}$) | $\delta^{25}Mg$ (‰) | $\delta^{26}Mg$ (‰) | $^{26}Al/^{27}Al$ | Al/Mg[2] |
|---|---|---|---|---|---|---|---|---|---|
| A04-04-#009 | 1 | spinel | 305 | 27.9±1.8 | 18.2±1.5 | 105±53 | −37±47 | N.D.[3] | 1.9 |
| A05-03-#028 | 1 | spinel | 200 | 20.2±3.4 | 14.3±3.1 | 49±49 | 22±47 | $<3.1\times10^{-3}$ | 1.9 |
| A05-07-#001 | 1 | spinel | 305 | 14.2±1.2 | 19.5±1.3 | 35±31 | −5±29 | $<1.6\times10^{-4}$ | 2.8 |
| A05-08-#020 | 1 | spinel | 305 | 10.0±1.5 | 16.6±2.0 | 24±21 | −11±20 | N.D. | 1.2 |
| A06-04-#018 | 1 | spinel | 305 | 10.0±1.7 | 27.3±2.5 | 144±61 | 65±56 | $<1.4\times10^{-3}$ | 2.0 |
| A06-09-#001 | 1 | spinel | 335 | 7.6±0.6 | 19.8±0.8 | 53±31 | −2±29 | N.D. | 3.9 |
| A09-02-#017 | 1 | spinel | 295 | 3.4±0.8 | 12.2±1.9 | −22±24 | 31±25 | $(2.0\pm0.1)\times10^{-3}$ | 2.4 |
| A09-04-#024 | 1 | spinel | 245 | 9.4±1.4 | 22.0±2.7 | 34±31 | 30±31 | $<1.9\times10^{-3}$ | 2.2 |
| A10-05-#013 | 1 | spinel | 230 | 11.3±1.8 | 19.5±2.6 | −5±47 | −2±46 | $(7.3\pm6.1)\times10^{-5}$ | 2.8 |
| A11-02-#001 | 1 | spinel | 305 | 11.1±1.6 | 21.6±2.2 | −138±59 | 8±62 | $(6.3\pm0.9)\times10^{-3}$ | 2.1 |
| A13-02-#002 | 1 | spinel | 255 | 7.2±0.8 | 12.2±1.3 | 43±27 | −14±26 | N.D. | 1.4 |
| A14-03-#002 | 1 | spinel | 255 | 13.0±1.8 | 24.8±2.5 | 41±28 | 50±28 | $(3.7\pm0.8)\times10^{-4}$ | 2.0 |
| A15-00-#019 | 1 | spinel | 290 | 19.6±2.2 | 21.5±2.4 | 115±53 | 355±69 | $(1.1\pm0.1)\times10^{-2}$ | 2.2 |
| A15-03-#012 | 1 | spinel | 230 | 11.3±1.6 | 22.9±2.4 | −32±46 | 28±46 | $(2.7\pm0.4)\times10^{-3}$ | 2.1 |
| A15-10-#010 | 1 | spinel | 280 | 9.2±1.4 | 13.0±2.1 | 72±33 | 38±31 | $<7.8\times10^{-4}$ | 1.9 |
| A15-11-#045 | 1 | spinel | 240 | 49.0±7.1 | 18.3±4.5 | −14±43 | 96±45 | $(5.2\pm0.6)\times10^{-3}$ | 2.0 |
| A16-06-#023 | 1 | spinel | 275 | 9.7±1.4 | 16.9±2.2 | 51±57 | 58±56 | $(2.6\pm1.6)\times10^{-4}$ | 2.2 |
| A20-03-#030 | 1 | spinel | 220 | 19.0±3.2 | 19.4±3.3 | −318±110 | −139±120 | $(1.3\pm0.4)\times10^{-2}$ | 1.3 |
| A21-01-#001 | 1 | spinel | 305 | 12.3±0.6 | 15.8±0.7 | 14±21 | 24±21 | $(5.7\pm0.5)\times10^{-4}$ | 1.5 |
| A10-07-#008 | 1 | corundum? | 255 | 14.7±2.0 | 12.8±2.3 | −111±96 | 1233±160 | $(7.4\pm0.6)\times10^{-4}$ | 154.4 |
| A11-04_#013 | 2 | spinel | 265 | 9.8±1.5 | 6.78±2.2 | 27±27 | 13±26 | $<1.2\times10^{-3}$ | 1.8 |
| A14-09-#027 | 2 | spinel | 240 | 7.5±2.0 | 8.19±3.0 | 40±45 | 29±43 | $<2.6\times10^{-3}$ | 2.4 |
| A16-07-#006 | 2 | spinel | 345 | 8.8±0.7 | 8.65±1.1 | 34±53 | 127±54 | $(4.2\pm0.6)\times10^{-3}$ | 1.5 |
| A19-03-#012 | 2 | spinel | 230 | 13.1±1.9 | 7.20±2.4 | 35±61 | −60±55 | N. D. | 3.9 |
| A15-08-#009 | nova | spinel | 240 | 319.8±11.4 | 15.6±2.3 | 70±46 | −25±42 | N. D. | 1.2 |

[1] The grain sizes were estimated based on $^{27}Al^{16}O^-$ ion images.
[2] The 1σ uncertainty is ±15% according to the SRM 610 measurements.
3 N. D. denotes that no $^{26}Mg$ excess after GCE correction was detected within 1σ uncertainties.

**Supplementary Material**

**Table A1.** Oxygen isotopic compositions of additional presolar oxides from Orgueil acid residue.

| Grain Name | Group | Diameter (nm) | $^{17}O/^{16}O$ ($\times 10^{-4}$) | $^{18}O/^{16}O$ (($\times 10^{-4}$) |
|---|---|---|---|---|
| A03-01-#015 | 1 | 370 | 11.5±1.0 | 16.4±1.1 |
| A03-03-#157 | 2 | 305 | 9.7±1.5 | 8.0±2.1 |
| A04-03-#037 | 4 | 125 | 2.6±0.3 | 19.5±0.7 |
| A05-02-#006 | 1 | 305 | 7.3±0.8 | 17.8±1.7 |
| A05-07-#09 | 1 | 305 | 8.9±1.2 | 18.8±2.0 |
| A06-02-#064 | 4 | 305 | 2.0±0.4 | 21.7±1.0 |
| A06-04-#086 | 4 | 400 | 2.4±0.3 | 20.3±0.9 |
| A06-07-#158 | 4 | 335 | 1.7±0.5 | 18.2±1.1 |
| A06-07-#169 | 4 | 470 | 2.2±0.3 | 20.9±0.7 |
| A06-07-#144[b] | 4 | 385 | 2.6±0.3 | 21.5±0.7 |
| A07-06-#053 | 1 | 245 | 16.5±3.2 | 24.7±3.9 |
| A07-08-#019 | 2 | 255 | 5.6±1.9 | 6.1±3.5 |
| A07-08-#002 | 1 | 265 | 3.2±0.7 | 11.5±1.7 |
| A08-00-#031 | 1 | 265 | 11.6±1.7 | 19.0±2.3 |
| A09-00-#014 | 2 | 255 | 5.5±1.0 | 9.1±2.0 |
| A09-01-#062[b] | 1 | 240 | 3.7±1.2 | 10.5±2.5 |
| A09-02-#278 | 2 | 265 | 6.2±2.0 | 8.2±3.0 |
| A09-02-#035 | 1 | 210 | 20.1±2.5 | 19.2±2.5 |
| A09-03-#080 | 1 | 640 | 4.0±0.1 | 22.8±0.4 |
| A09-04-#029 | 4 | 330 | 2.6±0.4 | 23.6±1.0 |
| A09-07-#145 | 4 | 670 | 2.0±0.5 | 22.1±1.3 |
| A10-02-#055[b] | 1 | 420 | 3.6±0.2 | 23.4±0.6 |
| A10-04-#013 | 1 | 245 | 11.4±1.9 | 19.5±2.5 |
| A10-08-#123 | 2 | 280 | 10.2±1.7 | 6.6±2.4 |
| A12-01-#052 | 1 | 430 | 3.6±0.2 | 22.9±0.5 |
| A14-01-#74 | 1 | 410 | 4.5±0.3 | 23.8±0.7 |
| A14-01-#094 | 1 | 255 | 13.2±2.8 | 12.2±3.4 |
| A15-00-#106 | 1 | 500 | 7.1±0.8 | 17.9±1.5 |
| A15-11-#049 | 1 | 255 | 14.7±2.6 | 23.6±3.7 |
| A15-08-#226 | 1 | 295 | 33.9±4.6 | 16.6±3.5 |
| A15-08-#213 | 1 | 445 | 34.5±3.4 | 21.1±2.6 |
| A15-08-#232 | 1 | 325 | 16.0±3.0 | 21.4±4.0 |
| A15-08-#104 | 1 | 345 | 10.5±1.6 | 22.5±2.6 |
| A15-08-#273 | 1 | 290 | 16.9±3.1 | 28.9±4.1 |
| A15-09-#079 | nova | 345 | 79.7±5.6 | 18.2±2.7 |
| A16-02-#077 | 3 | 275 | 2.8±1.0 | 10.3±2.4 |
| A16-02-#037 | 4 | 210 | 2.1±1.2 | 36.2±3.6 |
| A16-04-#056 | 1 | 245 | 15.6±2.6 | 21.8±3.1 |
| A16-05-#131 | 1 | 330 | 4.1±1.1 | 11.8±2.0 |
| A17-06-#165 | 4 | 570 | 2.6±0.7 | 14.5±1.6 |
| A17-09-#028 | 1 | 275 | 7.7±1.3 | 11.1±2.0 |
| A18-02-#265 | 4 | 280 | 1.8±1.0 | 12.2±2.2 |
| A18-02-#044 | 4 | 255 | 7.0±1.6 | 47.0±4.0 |
| A19-05-#097 | 1 | 730 | 6.2±0.6 | 18.9±1.1 |
| A20-02-#197 | 3 | 310 | 1.5±1.7 | 7.4±3.2 |
| A21-06-#062 | 4 | 510 | 2.7±0.2 | 21.2±0.5 |
| A21-08-#164 | 1 | 315 | 7.2±0.7 | 21.5±1.2 |
| A05-03-#021[a] | 4 | 310 | 1.4±0.9 | 54.6±6.2 |
| A05-03-#031[a] | 4 | 255 | 0.9±0.9 | 9.2±2.9 |

[a]: These two grains are Al-poor as denoted in Fig. 1a.

[b]: These three grains were detected as outliers in only three of the four cases shown in Fig. A4. See text in Presolar grain selection criterion for details.

## Presolar Grain Selection Criterion

Different criteria (e.g., 3σ, 4σ, 5.3σ O isotope anomalies) have been adopted in the literature for identifying presolar grains. For presolar oxide grains, the criterion of 3σ anomalies was often adopted in previous studies (*e.g.*, Nittler et al. 2008), as it was found that in the acid-resistant oxide residues of primitive meteorites ~2% of the remaining grains are presolar in origin (Zinner et al. 2003), significantly higher than the expected fraction of population outside of the 3σ range for a Gaussian distribution, 0.27%. However, when we applied the 3σ criterion to our Orgueil residue, the inferred presolar oxide abundance is only ~0.3%, comparable to the expected percentage of data in the tail of a Gaussian distribution. The distributions of the $\delta^{17}O$ and $\delta^{18}O$ anomalies of our automatically identified O-rich grains (~30,000; see Section 2 in the main text for details) on the Orgueil mount are shown in Fig. A1. The Gaussian distributions (red and blue dots) in Fig. A1 are fits to the O isotopic data of all the identified O-rich grains (grey bars).

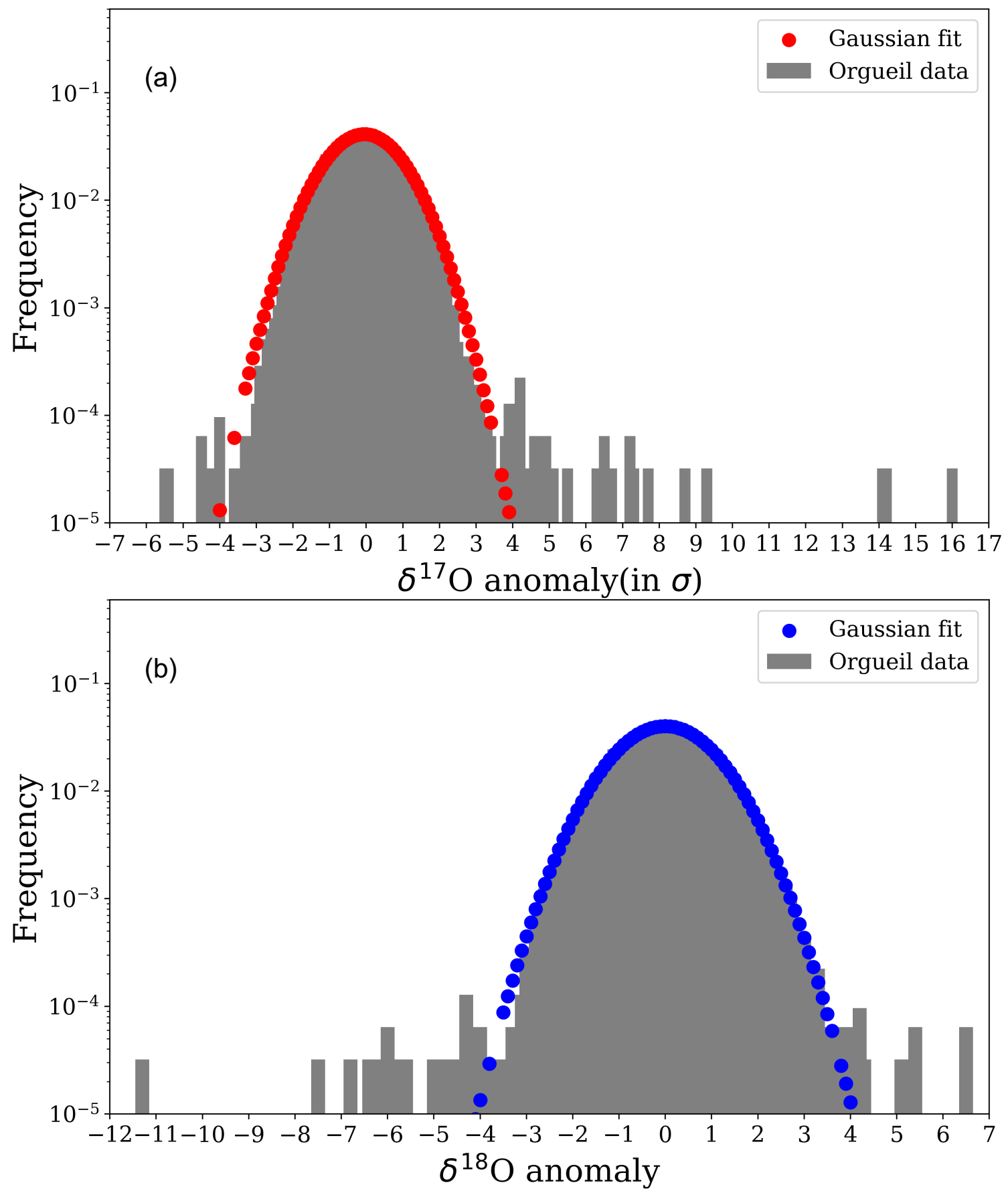

**Figure A1.** The probability distributions of $\delta^{17}O$ and $\delta^{18}O$ anomalies (in σ) of all O-rich particles present in the $^{16}O^{-}$ ion images from this study.

To flag anomalous grains, we have used the elliptic envelope outlier detection algorithm implemented in Python (EllipticEnvelope module in the scikit-learn 0.24.2 tool suite; Pedregosa et al. 2011). The algorithm detects outliers by assuming that the regular data come from a multivariate normal distribution (see Rousseeuw and Driessen 1999 for details). This algorithm works by first estimating the mean and covariance matrix of the dataset uncontaminated with outliers, to then flag the outlier data points as those that do not conform to this distribution. A difficulty is that one must provide as input an estimate of the fraction of outlier data expected (*contamination* percentage). We discuss below (*i*) how we tested the assumption that the joint $\delta^{17}O$-$\delta^{18}O$ distribution of non-presolar grains is a bivariate normal distribution and (*ii*) how we selected the *contamination* percentage used in the EllipticEnvelope module.

(*i*) To demonstrate that our analytical procedure results in a bivariation normal distribution for $\delta^{17}O$-$\delta^{18}O$, we used previously acquired NanoSIMS O isotopic imaging data for fine-grained chondritic materials embedded within a CM-clast (022-005) from the howardite Kapoeta (Liu et al. 2020). We chose this sample because we expect no detection of presolar O-rich grains (except for the Al-rich oxide 022-005-2 reported by Liu et al. 2020) in the dataset, the grains are comparable in size to the Orgueil O-rich grains studied here (*i.e.*, comparable sample properties), and the Kapoeta CM-clast and Orgueil O isotopic measurements were conducted under similar NanoSIMS analysis condition. For the CM-clast data reduction, we automatically defined ROIs using the "grid pattern-hexagons" option in L'image, and each ROI was defined to have a diameter of 170 nm. We complied a dataset consisting of O isotopic data for ~30,000 ROIs, like the number of O-rich grains found in this study, over six $10 \times 10$ μm areas. The bulk O isotopic compositions of the six areas were reported in Fig. 3 of Liu et al. (2020), and the bulk compositions of each of the areas were used for normalization for the identified ROIs within the respective areas. We show the distributions of the $\delta^{17}O$ and $\delta^{18}O$ anomalies of these CM-clast O-rich grains in Fig. A2 for comparison with the Orgueil O-rich grains in Fig. A1. The Gaussian fits to the CM clast data in Fig. A2 have means of −0.007 and 0.004 (compared to zero for a perfect Gaussian distribution), respectively, and standard deviations of 1.002 and 1.125 (compared to one for a perfect Gaussian distribution), respectively, thus demonstrating that the O isotopic data closely follow a Gaussian distribution when no presolar grain is present. This, therefore, justifies the Gaussian fits to the O-rich grain data from this study in Fig. A1 and the use of the ellipse envelope algorithm. The density distributions of the Orgueil O-rich grains are further illustrated in Fig. A3.

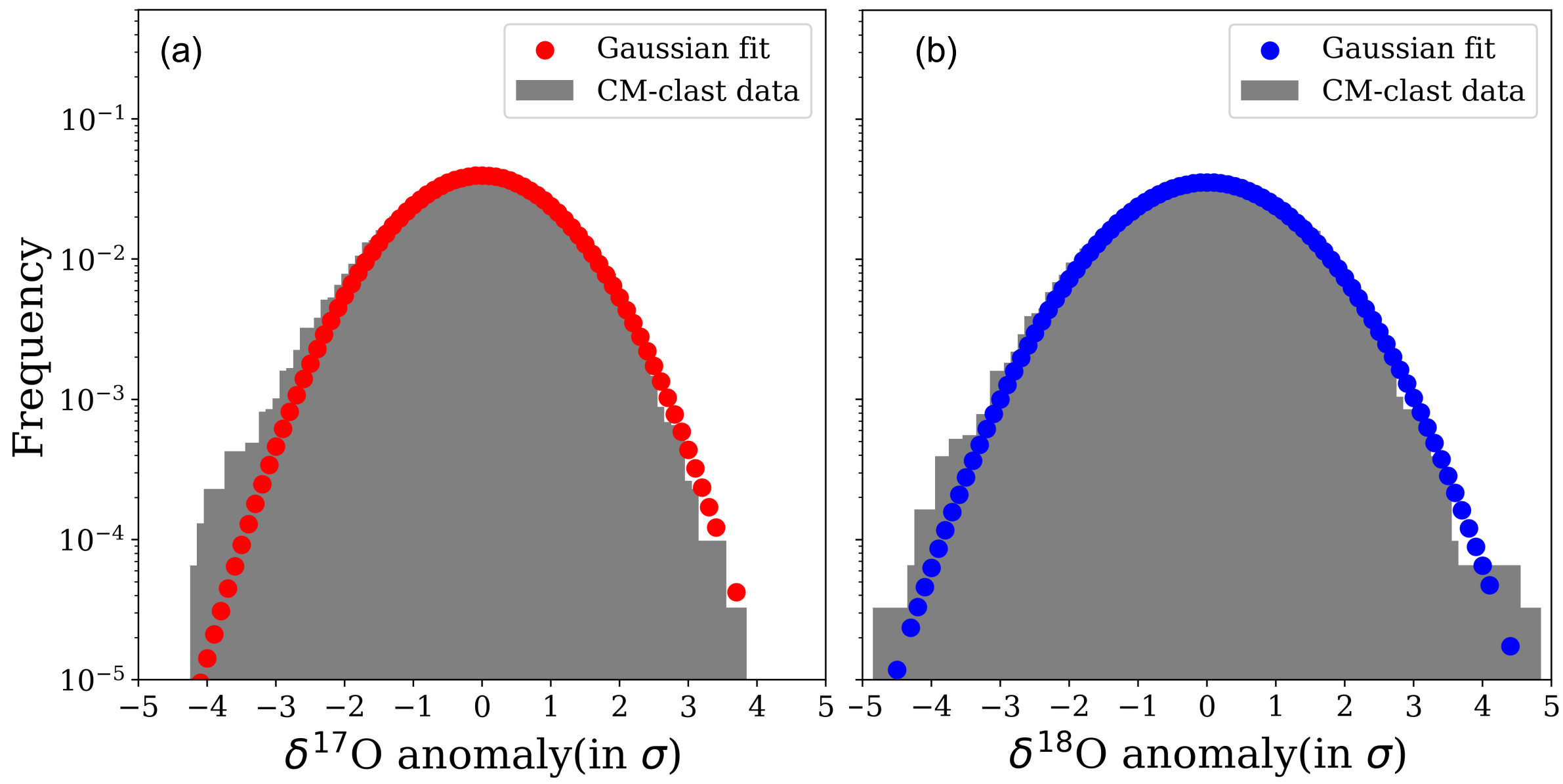

**Figure A2**. The same as Fig. A1, but for all O-rich particles embedded within a CM carbonaceous chondritic clast from Liu et al. (2020). See text for more details.

(*ii*)The only parameter in the EllipticEnvelope algorithm is *contamination*, which denotes the prior assumption for the percentage of outliers in the dataset. Since we know that we need to adopt the 4σ criterion for the univariate ($\delta^{17}$O or $\delta^{18}$O anomaly), we used the expected binormal distribution as a guidance and adjusted the parameter *contamination* until the majority of grains lying outside of the 4σ ellipsoid was recognized by the algorithm as outliers (Fig. A4); we wrote our own python code to visualize the outlier detection data and different σ ellipsoids (Fig. A4). In our visualization code the σ ellipsoids were enlarged by 13% (this number is specifically chosen for the 4σ ellipsoid and should be smaller for >4σ ellipsoids and larger for <4σ ellipsoids) by considering that the prescribed confidence level within a certain ellipsoid decreases with increasing dimensionality (Wang et al. 2015). The obtained number for *contamination* is 0.26%.

We then further tested the reliability of the EllipticEnvelope module results by employing another outlier detection algorithm in Python, LocalOutlierFactor (LOF) module in the scikit-learn 0.24.2 tool (Breunig et al. 2000). This algorithm was used recently by Ptáček et al. (2019) to flag spurious sediment chemical composition data. The algorithm measures the local deviation of density of a given sample with respect to its neighbors. The locality is given by *k*-nearest neighbors, whose distance is used to estimate the local density. By comparing the local density of a grain to the local densities of its neighbors, one can identify grains that have a substantially lower density than their neighbors, which correspond to outliers. Compared to the EllipticEnvelope algorithm, the LOF algorithm assumes no specific shape for the grain distribution but has an additional parameter, the number of neighbors (n_neighbors). We adopted 0.26% for the parameter *contamination* in the LOF module and analyzed our O-rich grain dataset by adopting different n_neighbors numbers (200, 500, 5000; Figs. A4b-c). A comparison of the LOF detection results shows that the detected outliers are almost independent of the choice of n_neighors except grains in the region highlighted by the red circle in Figs. A4b-c.

Figure A4 demonstrates a general consistency among the different sets of data. We detected a total of 81 outliers, and 75 of the grains were detected in all the four cases, three in three of the four cases (including two *Group* 1 grains and one *Group* 4 grain; highlighted in Table A1), and three in one case (removed from the list of presolar grains). Four of the 78 outliers have positive $\delta^{17}O$ and $\delta^{18}O$ anomalies but less than 150‰ and thus could be solar instead of presolar grains given similar O isotopic ratios observed in solar system objects (Sakamoto et al. 2007). These four grains are, therefore, also removed from our list of presolar grains, and the remaining 74 outliers are considered as presolar grains; their O isotopic compositions are summarized in Tables 1 and A1 and shown in Fig. 1. Since we could not analyze the literature data the same way as done for our Orgueil O-rich grains, we simply adopted the criterion of $\geq 4\sigma$ $\delta^{17}O$ or $\delta^{18}O$ anomalies that deviate from the terrestrial value by > 15% for selecting the literature grains for comparison with our O-rich grains in Fig. 1. The $4\sigma$ criterion is generally supported by our presolar grain data as 69 of the 74 grains have $\geq 4\sigma$ anomalies in $\delta^{17}O$ or $\delta^{18}O$.

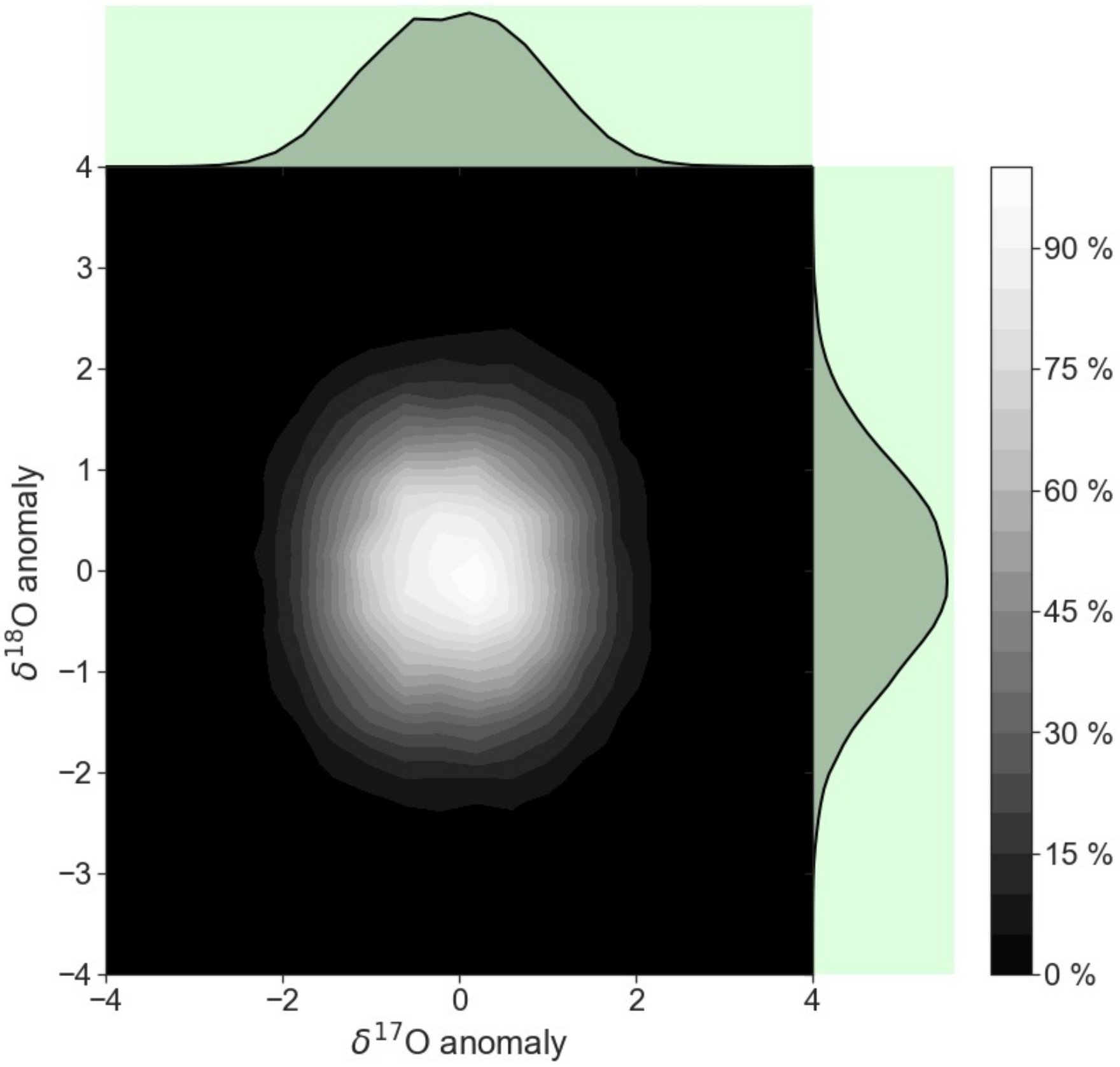

**Figure A3**. The density distribution of $\delta^{17}O$ and $\delta^{18}O$ anomalies (in $\sigma$) of the same set of data as in Figs. A1 and A2.

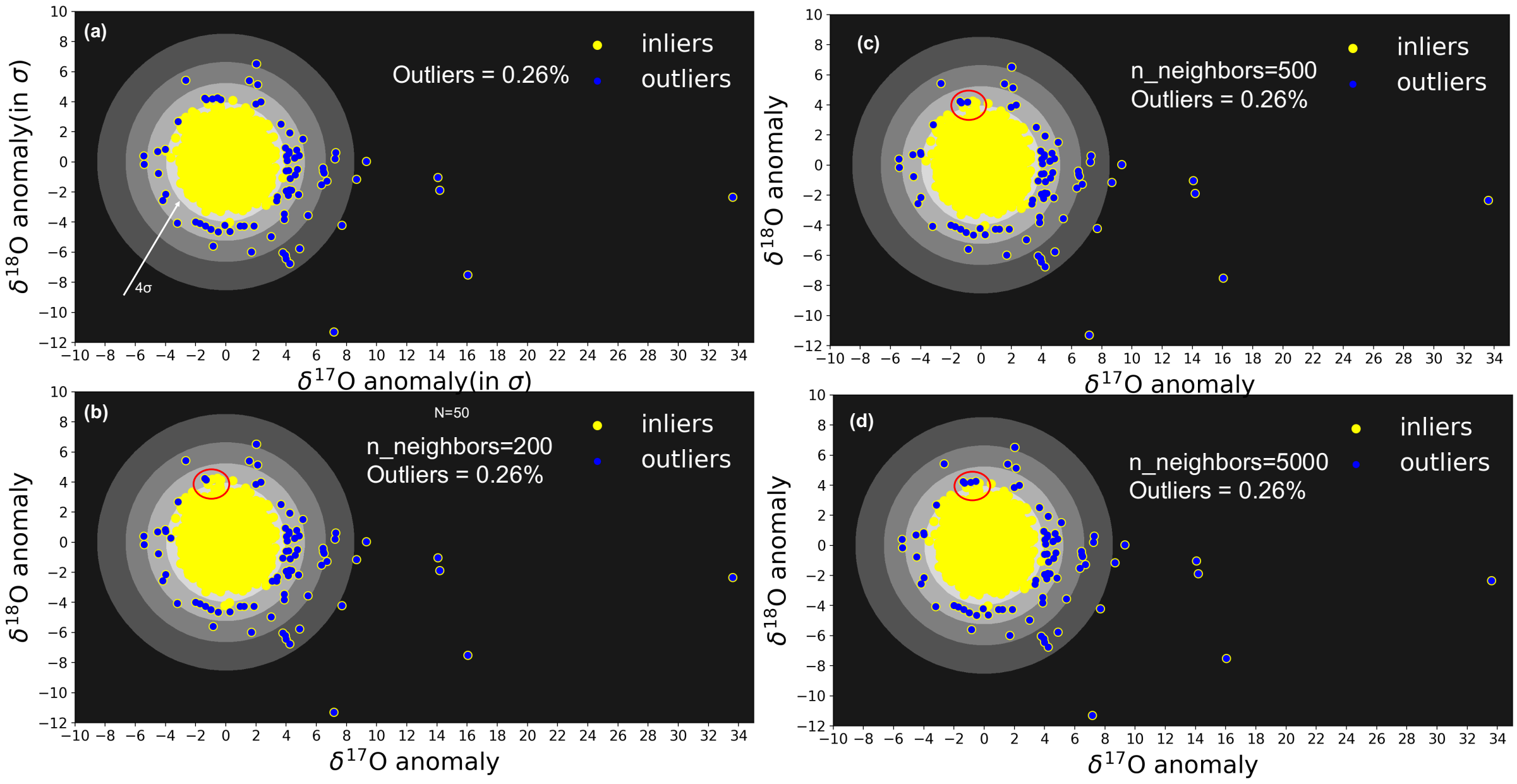

**Figure A4**. The distributions of $\delta^{17}O$ and $\delta^{18}O$ anomalies (in $\sigma$) of the same set of data as in Figs. A1-A3. The yellow dots represent grains that are detected as inliers and blue grains as outliers. A 2D Gaussian distribution is shown in the background (shown in grey colormap) for comparison with the grain data, and the $4\sigma$ ellipsoid is highlighted by a white arrow. Panel (a)

shows the EllipticEnvelope results and panels (b-c) the LOF results by adopting different n_neighbors numbers.